\newcommand{\non}{\nonumber}
\newcommand{\beq}{\begin{equation}}
\newcommand{\eeq}{\end{equation}}
\newcommand{\barr}{\begin{eqnarray}}
\newcommand{\earr}{\end{eqnarray}}
\newcommand{\la}{\langle}
\newcommand{\ra}{\rangle}
\newcommand{\gsim}{\mbox{\raisebox{-1.ex}
{$\stackrel{\textstyle >}{\textstyle\sim}$}}}
\newcommand{\lsim}{\mbox{\raisebox{-1.ex}
{$\stackrel{\textstyle <}{\textstyle \sim}$}}}
\newcommand{\square}{\kern1pt\vbox{\hrule height
1.2pt\hbox
{\vrule width1.2pt\hskip 3pt \vbox{\vskip 6pt}
\hskip 3pt\vrule width 0.6pt}\hrule
 height 0.6pt}\kern1pt}
\documentstyle[aps,prd]{revtex}

\begin{document}
\baselineskip = 14pt

\begin{titlepage}
\baselineskip .15in
\vskip 1.5cm
\begin{center}
{\large\bf Renormalization Group Approach in
Newtonian Cosmology}
\vskip .8in

Yasuhide {\sc Sota}$^{(a)}$, Toshiyuki  {\sc
Kobayashi} $^{(b)}$,
 Kei-ichi {\sc Maeda}$^{(c)}$,\\[.5em]
{\em Department of Physics, Waseda
University, 3-4-1 Okubo, Shinjuku-ku, Tokyo 169,
JAPAN}\\
\vspace{0.8cm} Tomomi {\sc Kurokawa}$^{(d)}$,
Masahiro {\sc  Morikawa}
$^{(e)}$,\\[.5em] {\em Department of Physics,
Ochanomizu University, 1-1 Otuka,2 Bunkyo-ku,
Tokyo 112, JAPAN}\\
\vspace{0.5cm} and\\
\vspace{0.7cm}
Akika {\sc Nakamichi}$^{(f)}$\\[.5em]
{\em Gunma Astronomical Observatory, 1-18-7
Ohtomo,  Maebashi,  Gunma 371, JAPAN} \\
\end{center}
\vfill
\begin{abstract}
We apply the renormalization group (RG) method to
examine  the  observable  scaling properties in
Newtonian cosmology. The original scaling
properties of the equations of motion in our
model are modified for averaged observables on 
constant  time slices. In the RG flow diagram, we
find three robust fixed points:
Einstein-de Sitter, Milne and Quiescent fixed
points. Their stability (or instability) property
 does not change under the effect of
fluctuations. Inspired by the inflationary
scenario in the early Universe, we set the
Einstein-de Sitter fixed point with small 
fluctuations, as the boundary condition at the
horizon scale. Solving the RG equations under
this boundary condition toward the smaller
scales, we find generic behavior of observables
such that the density parameter
$\Omega$ decreases, while the Hubble
parameter
$H$ increases for smaller averaging volume.  The
quantitative scaling properties are analyzed by
calculating the characteristic exponents around
each fixed point. Finally we argue the possible
fractal structure of the Universe beyond the
horizon scale.
\end{abstract}
\vfill
\vfill

\hskip 1.5cm (a) electronic mail :
sota@gravity.phys.waseda.ac.jp,

\hskip 1.5cm (b) electronic mail :
kobayasi@gravity.phys.waseda.ac.jp,

\hskip 1.5cm (c) electronic mail :
maeda@gravity.phys.waseda.ac.jp,

\hskip 1.5cm (d) electronic mail :
tk385@phys.ocha.ac.jp,

\hskip 1.5cm (e) electronic mail :
hiro@phys.ocha.ac.jp,

\hskip 1.5cm (f) electronic mail :
akika@astron.pref.gunma.jp
\end{titlepage}
\section{Introduction}
\par Several observational evidences imply scaling properties in the present 
Universe.
For example,
(a) G. de Vaucouleurs compiled data of
density-size relation for galaxies and clusters of
galaxies,   and pointed out the scaling relation of
$\log{\rho}=-1.7\log{L}$, where
$L$ is the linear scale of each astrophysical object
and $\rho$ is the mass density averaged at this scale
\cite{devaucouleurs70}. This relation asserts the
systematic decrease of the density at larger scale.
(b) It is widely known that the observed two-point
correlation function $\xi(r)$ for galaxies or for
clusters has the scaling property as $\xi(r)
\propto r^{-1.8}$
\cite{peebles80}.
(c) Pietronero and his collaborators
discuss a fractal structure of the Universe and claim
that the fractal dimension is about two up to the scale
of 1000 Mpc\cite{coleman92}.
(d) The observations of mass-luminosity ratio for the 
various astronomical objects
linearly increases toward the larger scale 
up to the scale about 10Mpc\cite{bahcall95}.
\par 
Motivated by the above observational evidences, 
we explore systematic underlying physics which may
govern the scaling properties of observables in the present Universe.
Usually the above scaling properties are understood as a result of an 
interplay of
the scale invariant initial condition (Harrison-Zel'dovich spectrum) and
the gravitational instability \cite{peebles80}.
In this usual approach, a global uniform background is assumed.
In our approach, we do {\it not} assume such uniform background a priori but 
consider general inhomogeneous distributions.
\par 
Let us consider the scaling properties carefully.
We start our analysis with the equations of motion for fluid in
the Newtonian Universe neglecting the pressure term for simplicity.
This set of equations admits a naive scale invariance; a
scaling of space-time coordinates as well as an appropriate scaling of 
observables leaves the set of equations unchanged.
We define this naive scaling transformation as the {\it operation-S}.
Such a naive scaling property is,
however, not necessarily reflected in the actual
observations in a direct form. This is because in
general the physically measurable quantities are strongly
affected by the actual method of observations.
Therefore the above naive scaling should be modified when we apply it
to the actual observations.
There are at least two important factors for the modification:
(1) An observable region is limited by causality.
(2) We may merely observe averaged quantities in a certain region in 
space.
\par We first consider the factor (1).
Actually, we can only observe a light signal from
a galaxy located on our past light-cone.
In Newtonian cosmology, the observable quantities are
located on  a  hyper-slice of constant time.
Because in general the above naive scale transformation
also requires a change of the time variable, we have
to adjust it so that we can
compare the observable quantities with different
scales on the same time slice.
This adjustment can be implemented by the time evolution of the observables 
by using the equations of motion for them.
We define this adjustment as the {\it operation-T}.
In this context, we have to
reconsider exactly what we are measuring. 
In this paper we consider scaling behavior of the cosmological parameters.
For example, the density parameter
$\Omega$ is a dimensionless quantity and therefore
it is  invariant under the naive scale transformation.
However in many cases,  we can directly measure the
matter density $\rho$ and separately the cosmic
expansion parameter $H$\cite{observation}. 
The combination of these observations {\it induces} the value of the density 
parameter $\Omega$ \cite{foot1} which depends on the scale.
 In this
case, the scale transformation and the necessary time
adjustment are non-trivial. 
Thus we have to carefully
distinguish the direct observables and the secondary
{\it induced} observables.
\par
Next we consider the factor (2).
What we observe is necessarily an averaged quantity
on some  region  of space-time because there are
always spatial fluctuations and the resolution of our
observation is limited. Therefore the observables are
most generally dependent on the scale of averaging.
We define this averaging as the {\it operation-A}.
However, it would be very difficult to consider the
invariant averaging procedure in the full general
relativity and to make connection with
observables\cite{averaging_GR}. Therefore mainly from
this reason we would better restrict our considerations
to the Newtonian cosmology, in which an averaging procedure
is properly defined and is explicitly
calculated\cite{buchert95}.
\par
After the above adjustments
(1) the operation-T and
(2) the operation-A
on the original naive scaling properties of the underlying
equations of motion, we would obtain the relevant information
for actually observable quantities.
Thus in general, the difference of two observables at
different scales would be related with each other by the naive scaling
(operation-S)
compensating time evolution (operation-T)
and the inevitable averaging procedure (operation-A).
\par
In order to unify the above operations, S, T, and A, we need to 
introduce one more ingredient in our considerations.
The most
sophisticated method so far to deal with such 
multiple operations
would be the renormalization group (RG)
method\cite{wilson74}.
This is a very general method to obtain the observable
response of the system against the scale change and
is widely used in various fields in physics including
quantum field theory, statistical mechanics, fluid
mechanics, etc.
For example in the quantum field theory, a naive scaling in the classical
theory should be modified by the quantum fluctuations.
In the 
statistical physics, 
the scaling property is used with
the averaging operation to yield 
critical exponents.

In this paper, we try to apply this RG method to Newtonian Cosmology.
In this model, the scaling invariance holds all the time.
Thanks to the averaging procedure, we can define 
classical fluctuations of the observables.
Then, naive scaling behaviors of the observables are modified by renormalization of the classical 
fluctuations.
This is conceptually the same as RG methods in other fields of physics.


\par 
The RG equations we obtain give the effective
scale change of  averaged observables, 
and the effect of spatial fluctuations on observables. The
flow diagram generated by the RG equations in the
parameter space  of observables represents how the
observables actually change under the scale
transformation. Each flow line represents a single
physical system, in which the averaged
variable at different scale corresponds to each point
on the flow. Thus the RG flow gives a whole set of
possible cosmological models. The fixed points, i.e.
the stagnation points in the flow diagram, of the RG
equations represent the possible ``uniform"
structure of observables, that is the averaged
variables do not depend on the scale, and
characterize diversities of the system.
\par There have been some applications of the RG method in
cosmology so  far \cite{Vega}.
The RG method is an effective tool to examine the
asymptotic  behavior  of a dynamical system.
In fact, Koike, Hara and Adachi succeeded in
explaining the  critical behavior of a self-similar
solution during the gravitational  collapse of a
radiation fluid using the RG method
\cite{hara95}. Subsequent series of studies
 are in this line of
study\cite{iguchi97}. In  their cases, the
aspects of (S) and (T) are reflected in their method.
Thus their RG method does not necessarily contain an
averaging procedure (A).
On the other hand,
Carfora and Marzuoli\cite{carfora84} and Carfora and
Piotrkowska\cite{carfora95} explicitly accounted
 for the spatial fluctuation of physical quantities
in general relativity and derived the RG equation
  reflecting the aspects (S) and (A). Since they
concentrated on the scaling in the spatial direction,
   the aspect (T) has not been manifestly reflected
in  their approach. Here in this paper, to examine  the
scale dependence of the observables on a constant
hyper-slice, we apply the
   RG method from the full aspects of (S),(T), and
(A) altogether, assuming some scaling property in
averaging observables.
\par This paper is constructed as
follows: In section II, we derive the RG equation
for the averaged Newtonian  fluid. The section III
is the main part of our analysis. Here we  examine
the RG flow in the parameter space and show that
there appear three robust fixed points including the
Einstein-de Sitter universe. We examine the effects
of shear, tidal force, and fluctuations, and derive
the scale dependence of the physical variables such
as $\Omega$ and $H$ around the Einstein-de Sitter
universe. In section IV, we summarize our results.
We  also present the full RG equations including
fluctuations up to
 the second order cumulants in the Appendix.

\section{Derivation of Renormalization Group Equations}
In
Newtonian cosmology, the following equations
describe the  evolution of a self-gravitating fluid
 in the Eulerian coordinate system:
\barr
{\partial \rho\over \partial t} &=& - \nabla \cdot
(\rho
\mbox{\boldmath$v$}) \ ,
\non\\
{\partial \mbox{\boldmath$v$} \over \partial t} &=&
-   (\mbox{\boldmath$v$}
\cdot \nabla)\mbox{\boldmath$v$} +\mbox{\boldmath$g$}
\ , \non\\
\nabla\cdot\mbox{\boldmath$g$} &=& -4\pi G \rho \ ,
\label{fluid}
\earr
where $\rho$, $\mbox{\boldmath$v$}$, and
$\mbox{\boldmath$g$}~  (\equiv -\nabla\phi)$
respectively denote the mass density, the velocity
of the fluid,  and the gravitational acceleration.
Now we transform these equations into the Lagrangian
coordinate  system. In order to eliminate the
gravitational acceleration
$\mbox{\boldmath$g$}$, we introduce the spatial
derivative of   the velocity field and decompose it
into a trace part (expansion
$\theta = \nabla
\cdot
\mbox{\boldmath$v$}$), a symmetric traceless part
(shear
$\sigma^i_{~j}$)  and an anti-symmetric part
(rotation
$\omega^i_{~j}$) (\cite{buchert95}).  Then the
above equations become
\barr
\frac{d \rho}{dt} &=& - \rho \theta , \label{frho}
\\
%
\frac{d \theta}{dt} &=&
- \frac{1}{3} \theta^2
+2\,( \omega^2 - \sigma^2 ) - 4 \pi G \rho ,
\label{fthet}\\
%
\frac{d \sigma^i_{~j}}{dt} &=&
-  \frac{2}{3} \theta \sigma^i_{~j} - \sigma^i_{~k}
\sigma^k_{~j} - \omega^i_{~k}
\omega^k_{~j} +\:\frac{2}{3}\,( \sigma^2 - \omega^2 )
\delta^i_{~j} - E^i_{~j} , \label{fsig}
\\
%
\frac{d \omega^i_{~j}}{dt} &=&
-  \frac{2}{3} \theta \omega^i_{~j}
-  \sigma^i_{~k} \omega^k_{~j} - \omega^i_{~k}
\sigma^k_{~j}
\label{fomg} ,
\earr
where the Lagrange full time-derivative
\beq
\frac{d}{dt} \equiv {\partial \over \partial t} +
\mbox{\boldmath$v$}
\cdot
\nabla \
\eeq
is used on the left hand side.
In these equations, $\sigma$ and $\omega$ denote the
magnitudes  of  shear and rotation, respectively,
defined by
\beq
\sigma^2={1 \over 2} \sigma^i_{~j}\sigma_i^{~j}
~~~{\rm and}  ~~~~\omega^2= {1 \over 2}
\omega^i_{~j}\omega_i^{~j} .
\eeq
The tensor
$E^i_{~j}=
\nabla^i\nabla_j \phi-\delta^i_{~j}\nabla^2 \phi/3
$  denotes the tidal force which comes from the
spatial  difference of the gravitational
acceleration.  In pure Newtonian dynamics, this is
a slaving variable and there is no equation  of
motion for it.  Fortunately, several excellent
approximation schemes have been developed to deal
with this  highly non-local quantity. In this
paper, we mainly follow the local tidal
approximation (LTA)
\cite{hamilton94}\cite{hamilton-foot}, in which the equation of motion
for the tensor
$E^i_{~j}$ is given by \barr
\frac{dE^i_{~j}}{dt} &=& -\theta \,E^i_{~j} - 4\pi
G\,\rho\,\sigma^i_{~j}.
\label{tidal}
\earr
Hereafter we neglect the rotation for simplicity
($\omega^i_{~j}=0$ ). Actually the  initial
assumption of
$\omega^i_{~j}=0$ is sufficient to guarantee   this
condition all the time as is seen from the above
equation. Thus irrotational motions  form a closed
subclass of the full dynamics. Then, we can
simultaneously diagonalize  both the matrices
$\sigma^i_{~j}$ and $E^i_{~j}$. We work in the frame
where they are diagonalized as $\sigma_i$ and $E_i$
($i = 1,2, 3$), respectively.  Since
$\sigma^i_{~j}$ and $E^i_{~j}$ are traceless,   only
the two components of each are independent. Then it
may be sometimes more convenient to introduce the
following variables $\sigma_\pm$ and
$E_\pm$:
\barr
(\sigma_1, \sigma_2, \sigma_3) &=&(\sigma_+
+\sqrt{3}\sigma_-,
\sigma_+-\sqrt{3}\sigma_-, -2\sigma_+) , \non\\
(E_1, E_2, E_3) &=&(E_++\sqrt{3}E_-,
E_+-\sqrt{3}E_-, -2E_+).
\earr

At a glance, we observe a scaling property in the
above set  of equations of motion (\ref{frho}) -
(\ref{fomg}),  (\ref{tidal}). They  are
invariant under the following scaling transformation:
\barr
t & \rightarrow & t' \equiv e^{s{\mit
\Delta}\tau} t \non\\
\mbox{\boldmath$x$} & \rightarrow &
\mbox{\boldmath$x$} ' \equiv e^{{\mit \Delta}\tau}
\mbox{\boldmath$x$}
\non\\
\theta & \rightarrow & \theta ' \equiv
e^{-s{\mit \Delta}\tau} \theta
\non\\
\sigma_\pm & \rightarrow & \sigma_\pm '  \equiv
e^{-s{\mit \Delta}\tau}
\sigma_\pm
\non\\
E_\pm & \rightarrow & E_\pm ' \equiv e^{-s{\mit
\Delta}\tau} E_\pm
\non\\
\rho & \rightarrow & \rho' \equiv e^{-2s{\mit
\Delta}\tau}
\rho,
\label{fluscal}
\earr
where $s$ is a constant free parameter and
${\mit \Delta}\tau =
\tau'-\tau$ measures a  change of scale.
\par
Now, based on these scaling properties, we derive
renormalization group equations for  spatially
averaged observables on a constant time slice. To
analyze the  structure of the Universe, we adopt a
scaling solution of our basic equations.  Such a
scaling solution could be realized as a result of a
self-gravitating dynamical system with a scale
invariant density fluctuation\cite{peebles80},
although here we do not specify any models to give
such a scaling solution.  If, however, we consider a
scaling solution as (\ref{fluscal}) as it is,  we
find either a homogeneous Universe or inhomogeneous
Universe with a "center",  which is not consistent
with the cosmological principle.  We are interested
in the inhomogeneous Universe model  consistent with
the cosmological principle such as a fractal
universe.  If we have a fractal universe, we expect
that some averaged values  around an observer may
show a scaling property. Such a scaling property
will be always observed anywhere because of a
fractal structure of the Universe,  which is
consistent with the cosmological principle. This is
why we  consider spatially averaged observables
here. We assume a scaling property  as
(\ref{fluscal}) not only for  averaged values of the
variables in the above equations but also for those
of the fluctuations such as a second order cumulant.

We define a spatial average $\langle ... \rangle  $
of an  observable $f(t,\mbox{\boldmath$x$} )$  at
time $t$  as
\barr
\la f{\ra}_{{\cal D} (t)} = \frac{1}{V(t)}
\int_{{\cal D} (t)}  d^3 x f(t,\mbox{\boldmath$x$} ),
\label{average}
\earr
where ${\cal D} (t)$ is a spatial domain at time $t$
and $V(t)$  is its volume defined by $
V(t)=\int_{{\cal D} (t)} d^3 x $. The observable
$f(t,\mbox{\boldmath$x$} )$ can be any function of
$\theta,
\rho,\sigma_\pm, E_\pm$.
We would like to derive the expression for the
infinitesimal  scale  change of the averaged
observable on the same time slice, i.e.
\barr
\frac{{\mit \Delta} \la f\ra_{{\cal D} (t)}}{{\mit
\Delta} {\ell }}  =\frac{\la  f\ra_{{\cal
D'} (t)}-\la f\ra_{{\cal D} (t)}}{{\mit \Delta}
{\ell }},
\label{eqofrg}
\earr
where the parameter ${\ell }$ measures the true
scale change on  this  time slice, which is defined
by
\barr
e^{{\mit \Delta} {\ell }} \equiv e^{\ell  ' -\ell
}=\left(
\frac{V'(t)}{V(t)} \right)^{1/3},
\label{truescalechange}
\earr
and is different from the naive scaling parameter
$\tau$. ${\cal  D'}(t)$ and $V'(t)$ denote the
domain and volume transformed back to the original
time $t$-slice by time  evolution from
${\cal D'}(t')$ and $V'(t')$,  which are obtained by
a scale transformation (\ref{fluscal}).  The above
equation (\ref{eqofrg}) can be decomposed into two
parts:\\ (i) a change
${\mit \Delta}_{S} \la f\ra $ associated with a
scale  transformation  from ${\cal D} (t)$ to ${\cal
D'} (t')$, \\ (ii) a change
${\mit \Delta}_{T} \la f\ra $ associated  with  a
time  evolution from
${\cal D'} (t')$ to ${\cal D'} (t)$. \\ That is,
\barr
\frac{{\mit \Delta} \la f\ra_{{\cal D} (t)}}{{\mit
\Delta} {\ell }}  &=&\frac{1}{{\mit \Delta} {\ell }}
\left[
\left( \la f\ra_{{\cal D'} (t')}-\la f\ra_{{\cal D}
(t)} \right)+
\left( \la f\ra_{{\cal D'} (t)}-\la f\ra_{{\cal D'}
(t')} \right)
\right]
\non\\
&\equiv &\frac{1}{{\mit \Delta} {\ell }} \left(
{\mit \Delta}_{S}
\la f\ra  +
{\mit \Delta}_{T}
\la f\ra \right).
\label{bunkai}
\earr
\par
As for part (i), we use the scaling law of the
observable $\la  f\ra_{{\cal D} (t)} $, which is
\begin{equation}
\la f\ra_{{\cal D} (t)} \rightarrow \la
f\ra_{{\cal D'} (t')} = e^{\alpha s {\mit \Delta}
\tau}
\la f\ra_{{\cal  D} (t)}.
\label{scalingf}
\end{equation}
\par
The second part (ii) is given by the Taylor
expansion of
$f(t',\mbox{\boldmath$x$}')$ and ${\cal D'} (t')$
with respect to time.
\barr
\la f \ra_{{\cal D'} (t')}
&=&\frac{1}{(1+ \la \theta\ra_{{\cal D'}(t)}  {\mit
\Delta} t)V'(t)}
\int_{{\cal D'} (t')} (1+ \theta {\mit \Delta} t) d^3
x'
\left[ f(t,
\mbox{\boldmath$x$}') +{df \over
dt}(t,\mbox{\boldmath$x$}'){\mit \Delta} t\right]
+O(({\mit \Delta} t)^2) \non
\\ &=&\la f \ra_{{\cal D'}(t)}+{\mit \Delta} t ~
\left(-\la \theta\ra_{{\cal D'}(t)} \la f
\ra_{{\cal D'}(t)}+
\la \theta f \ra_{{\cal D'}(t)} +\bigg< {df
\over dt}
\bigg>_{{\cal D'}(t)}
\right)+O(({\mit \Delta} t)^2),
\label{calte}
\earr
where ${\mit \Delta} t=t'-t$.
\par
The true scale change ${\mit \Delta} \ell $ on a
constant  time slice  is given as follows.  In the
scale transformation (i), the volume change is
$V'(t')/V(t)=e^{3{\mit \Delta} \tau}=1+3 {\mit \Delta}
\tau
$,  but it includes the expansion effect of the
Universe  because  time is also transformed.  To
obtain the true scale change on the same time slice,
we have to transform it back to the original time
slice by the time evolution (ii),  which gives
$V'(t')/V'(t)=1+\la \theta \ra {\mit \Delta} t $.
Then we  have
\barr
\left( \frac{V'(t)}{V(t)} \right)^{1/3}
&=&1+{\mit \Delta} \tau -\frac{\la \theta
\ra}{3}{\mit \Delta} t
\non  \\ &=&1+{\mit \Delta} \ell .
\earr
Thus we obtain
\barr
{\mit \Delta} \ell  = {\mit \Delta} \tau -\frac{\la
\theta
\ra}{3}{\mit \Delta} t.
\label{hi}
\earr
The time interval ${\mit \Delta} t = t'-t$ is given
by the  scale  transformation (\ref{fluscal}) as
\barr
\frac{{\mit \Delta} t}{{\mit \Delta}\tau}=s t .
\label{scaletimerel}
\earr
Hereafter we identify this fixed time $t$ as  the
present cosmic  time $t_0$, when we observe
cosmological quantities.
\par
Thus, combining Eqs.(\ref{bunkai}) $\sim$
(\ref{scaletimerel}),  we obtain the differential
equation  for the averaged observable
$\la f\ra_{{\cal D} (t)}$:
\barr
\frac{d \la f\ra_{{\cal D} (t)}}{d {\ell }}
&=&{\cal S} \left(  {\alpha \over t_0} \la
f\ra_{{\cal D} (t)} + \la \theta
\ra_{{\cal D} (t)}\la f\ra_{{\cal D} (t)} -\la \theta
 f \ra_{{\cal D} (t)}-\bigg< {df \over
dt}\bigg>_{{\cal D} (t)}
\right),
\label{rgeq}
\earr
where
\barr
{\cal S} =\frac{s}{1-s \la \theta \ra_{{\cal D} (t)}
t_0/3} .
\label{maeda}
\earr
The parameter $s$-dependence appears solely through
this factor
${\cal S}$.
\par
The equation (\ref{rgeq}) is not yet  sufficient to
obtain a  closed set of RG equations  because of the
non-linearity of the basic equations. Differential
equations for the $n$-th order averaged  quantities
are not closed up to the $n$-th order. They
necessarily contain $(n+1)$-th order averaged
quantities. This is the famous BBGKY hierarchy in
statistical physics. The ordinary method to  obtain
the closed set of equations is to truncate this
hierarchy at some order. In our case, we include the
effect of fluctuations at the lowest non-trivial
level, i.e. the second order cumulants. Then we will
neglect   all  intrinsic fluctuations (cumulants)
higher than the second order,  finding the following
truncation formula:
\beq
\langle f g h \rangle \rightarrow \langle f \rangle
\langle g h
\rangle + \langle g \rangle \langle f h \rangle +
\langle h
\rangle \langle f g \rangle - 2
\langle f \rangle \langle g \rangle \langle h
\rangle \, .
\eeq
To write down the basic equations, we shall use
the second order cumulant $\langle f  g
\rangle_{\rm c}$ instead of the averaged quadratic
quantity $\langle f  g
\rangle$, where $\langle f  g
\rangle_{\rm c}\equiv \langle f  g
\rangle - \langle f \rangle \langle g \rangle$.
Applying
this reduction, we obtain a closed set of 27
differential equations for averaged variables and
their   second order cumulants. We will present the
complete expression   in the Appendix. Here we
only show a reduced set of equations to the first
order (the averaged variables) by setting $\langle
f  g
\rangle_{\rm c}=0$:
\begin{eqnarray}
\frac{d \langle \rho \rangle}{d{\ell}}&=&
{\cal S}  \bigg[
\langle \theta \rangle - 2 \bigg] \langle \rho
\rangle,
\label{rerho}
\\
\frac{d \langle \theta \rangle}{d{\ell}}
&=&
{\cal S}  \bigg[
\left(
\frac{\langle \theta \rangle }{3} - 1 \right)
\langle \theta
\rangle  + 4\pi G \langle \rho \rangle + 6 (
\langle \sigma_+
\rangle^2 +\langle \sigma_- \rangle^2)
\bigg] ,
\label{retheta}
\\
\frac{d \langle \sigma_+ \rangle }{d{\ell}}&=&
{\cal S}  \bigg[
\left(\frac{2}{3} \langle \theta \rangle
 - 1 \right) \langle \sigma_+ \rangle
- \langle \sigma_+ \rangle^2
+ \langle \sigma_- \rangle^2
+ \langle E_+ \rangle\bigg],
\label{resig+}
\\
\frac{d \langle \sigma_- \rangle }{d{\ell}}&=&
{\cal S}  \bigg[
\left(\frac{2}{3} \langle \theta \rangle
 - 1 \right) \langle \sigma_- \rangle+ 2 \langle
\sigma_+
\rangle\langle \sigma_- \rangle  + \langle E_-
\rangle\bigg],
\label{resig-}
\\
\frac{d \langle E_+ \rangle}{d{\ell}}&=&
{\cal S}  \bigg[ (\langle \theta \rangle
- 2 ) \langle E_+ \rangle +
4\pi G \langle \rho \rangle \, \langle \sigma_+
\rangle
\bigg] , \label{retidn+}
\\
\frac{d \langle E_- \rangle}{d{\ell}}&=&
{\cal S}  \bigg[(\langle \theta \rangle -
2) \langle E_- \rangle +
4\pi G \langle \rho \rangle \, \langle \sigma_-
\rangle
\bigg].
\label{retidn-}
\end{eqnarray}
These also form a closed set of  equations at this
order.  We call these the first order RG equations,
while those in the Appendix the second order RG equations.

Here and
in Appendix,  we have set $t_0=1$ and dropped  the
suffix of integration domain for averaged values just
for simplicity.  Hence, in this units,  the expansion
$\theta$ and the density
$\rho$ for Einstein-de Sitter universe turn out to
be
$\theta_{\rm EdS}=2$ and
$\rho_{\rm EdS}=1/6\pi G $

\section{Analysis of RG Equations : Fixed Points and RG Flow}
%
We now examine the RG equations obtained in \S . 2.
 First, we analyze the first order RG
equations  (\ref{rerho})
$\sim$ (\ref{retidn-}). Although this may not be
realistic  because the structure of the Universe could be
highly inhomogeneous in our model, it
 shows a naive idea to understand the RG flow and
give  three important fixed points (including the
Einstein-de Sitter spacetime), which will be also most
relevant even in the case with fluctuations. We then
examine the stability of these fixed
points\cite{foot3}.  Next, introducing
fluctuations, we study the second order RG equations
(\ref{re2rho})
 $\sim$ (\ref{re2sig-e+}) and analyze the
stability of those fixed points  in a much wider parameter
space. Finally in this section,
 we set the Einstein-de Sitter  fixed point with small
amount of  fluctuations, as the  boundary condition at
the horizon scale in our present Universe and examine a
scale dependence of averaged variables   toward
smaller scale.
\subsection{Fixed points of the RG equations and
Stability }
First we analyze the structure of our RG equations.
From Eqs. (\ref{re2rho})
 $\sim$ (\ref{re2sig-e+}), we find that those  are
written in a vector form as
\barr
\frac{d\la F\ra }{d{\ell }}&=&{\cal S}\left(
A[\la F\ra ] + C \cdot \la FF\ra_{\rm c}\right)
\label{RGeq1}\\
\frac{d\la FF\ra_{\rm c} }{d{\ell }}&=&{\cal S} ~
B[\la F\ra ]\cdot\la FF\ra_{\rm c},
\label{RGeq2}
\earr
where $\la F\ra $ and $ \la
FF\ra_{\rm c}$ are 6- and 21-dimensional vectors
defined as $\la F\ra \equiv (\la f_i\ra)$ and
$\la FF\ra_{\rm c} \equiv (\la f_i f_j\ra_{\rm c})$ with
$f_i$'s ($i=1 \sim 6$) denoting $\rho ,
\theta  ,\sigma_+, \sigma_-, E_+$ and $E_-$.
$A[\la F\ra ]$, $B[\la F\ra ]$, and $C$ are
6-dimensional vector of quadratic functions of $\la
f_i\ra$, 21$\times$21 matrix of linear functions of
$\la f_i\ra$, and 6$\times$21 constant matrix,
respectively. Setting $\la FF\ra_{\rm c} =0$, we
have
\beq
\frac{d\la F\ra }{d{\ell }}={\cal S}
A[\la F\ra ] ,
\label{1st-RGeq}
\eeq
which is exactly  the first order RG
equations (\ref{rerho})
$\sim$ (\ref{retidn-}).

To
analyze the structure of the RG equations, we  first
need  to know the fixed point $(\la F\ra ,~\la
FF\ra )=(\la F\ra^*,~\la
FF\ra_{\rm c}^*)$, which is defined by
\beq
\frac{d\la F\ra }{d{\ell }}\bigg|_{\la F\ra =\la F\ra^*,
\la FF\ra_{\rm c}=\la FF\ra_{\rm c}^*}=
\frac{d\la FF\ra_{\rm c} }{d{\ell }} \bigg|_{\la F\ra =
\la F\ra^*,
\la FF\ra_{\rm c}=\la
FF\ra_{\rm c}^*}=0.
\eeq
As for the first order RG equations, the fixed point
$\la F\ra =\la F\ra^*$ is given by
\beq
A[\la F\ra^* ]=0.
\label{fixed_1st}
\eeq
The above equations (\ref{RGeq1}),({\ref{RGeq2})
with (\ref{fixed_1st}) guarantee that the fixed
points $\la F\ra^*$ in the first order RG
equations  are always   those for the second
order RG equations with
$\la FF\ra_{\rm c}^*=0$, although new
additional fixed points appear in the second order.

The stability analysis is  very important
to know asymptotic behaviors near the fixed points.
The stability is examined by linearizing the
RG  equations around the fixed point $\la F\ra^*,~\la
FF\ra_{\rm c}^*$ as
\barr
\la F\ra & = &\la F\ra^* +{\cal
F},
\\
\la FF\ra_{\rm c}&  =& \la FF\ra_{\rm c}^*  + ({\cal
FF})_{\rm c},
\earr
where both ${\cal F}$ and $({\cal FF})_{\rm c}$ are small
enough to make the linear perturbation treatment
effective around $\la F\ra^*,~\la
FF\ra_{\rm c}^*$
Then, we find the perturbation
equations in the form
\beq
 \frac{d}{d \ell }\left(\begin{array}{c}
 {\cal F}
\\[.5em]
 ({\cal FF})_{\rm
c}
\end{array}\right) = \left( \begin{array}{cc}
 {\cal W}_{11}  & {\cal W}_{12}
\\[.5em]
  {\cal W}_{21} &
 {\cal W}_{22}
\end{array}
\right)
\left(\begin{array}{c}
 {\cal F}
\\[.5em]
 ({\cal FF})_{\rm
c} ,
\end{array}\right)
\eeq
where  the perturbation matrix ${\cal
W}^{[2]}=({\cal W}_{AB})$ is given as
\beq
\left( \begin{array}{cc}
 {\cal W}_{11}  & {\cal W}_{12}
\\[.5em]
  {\cal W}_{21} &
 {\cal W}_{22}
\end{array}
\right)
=\left(\begin{array}{ccc}
{\partial ({\cal S}
A[\la F\ra ]) \over \partial\la F\ra }\bigg|_{\la F\ra
=\la F\ra^*} &~~ &
 {\cal S}^* C
\\[.5em]
{\partial ({\cal S}
B[\la F\ra ]) \over
\partial\la F\ra }\bigg|_{\la F\ra
=\la F\ra^*}\cdot\la FF\ra_{\rm c}^* &~~ &
{\cal S}^*
B[\la F\ra^*]
\end{array}
\right)
\eeq
with  ${\cal S}^*={\cal S}|_{\la F\ra
=\la F\ra^*}$.
For the first order RG equations, it turns out
to be
\beq
 \frac{d {\cal F}
}{d \ell }
 =
 {\cal W}^{[1]}
\label{1st_perturb}
\eeq
where  ${\cal W}^{[1]}$ is given as
\beq
 {\cal W}^{[1]}  = {\cal S}^*{\partial
A[\la F\ra ] \over \partial\la F\ra } \bigg|_{\la
F\ra  =\la F\ra^*} ,
\eeq
because of Eq. (\ref{fixed_1st})
For the fixed points in the first order RG equations,
we find the following important results.
As we mentioned, they are also the fixed points in
the second order with $\la
FF\ra_{\rm c}^*=0$. Then the perturbation matrix
${\cal W}^{[2]}$ becomes
\beq
 {\cal W}^{[2]}
=\left(\begin{array}{ccc}
 {\cal W}^{[1]} &~~ &
 {\cal S}^*~ C
\\[.5em]
0 &~~ &
{\cal S}^*~
B[\la F\ra^*]
\end{array}
\right) .
\label{2nd-matrix}
\eeq
This form guarantees that the eigenvalues of the first
order matrix ${\cal W}^{[1]}$ are always those of the
second order matrix ${\cal W}^{[2]}$.  As we will
show later, the other eigenvalues of the second
order matrix are constructed by a sum of any
two eigenvalues of the first order matrix (see
Eqs. (\ref{1st-2nd_1}), (\ref{1st-2nd_2})).

 The positive (negative) eigenvalue of  the
matrix
${\cal W}^{[2]}$ (or
${\cal W}^{[1]}$) shows the  instability  (stability) of
the fixed point
$(\la F\ra^*,\la FF\ra_{\rm c}^* $) (or $\la F\ra^*$)
toward the larger scale, i.e. in the large
$\ell$-direction.  The stability  and the instability
are exchanged  if we follow the RG flow in the opposite
direction, i.e.  toward the smaller scale.

The
eigenvalues  fully characterize the scale dependence of
all physical  variables in the vicinity of each fixed
point. Although
each observable is scaled according to its dimension,
the effect of dynamics changes the scaling property  from
the original bare scaling  law (\ref{fluscal}), so
that the scaling of each observable is determined by
the superposition of  several eigen functions as
follows:
\barr
\la f_i\ra  &=& \la f_i\ra_* + \sum_{m=1}^{N}
\epsilon_{i, m}  e^{\lambda_m \ell }
 \non\\
& &
\lambda_{\rm
min}=\lambda_1\leq\lambda_2\leq
\cdot \cdot \cdot \leq \lambda_N
=\lambda_{\rm max},
\label{eigen1}
\earr
where $\la f_i\ra_*$ is the value of
$\la f_i\ra$ at a fixed point, $\epsilon_{i,
m}$ is some small constant, and $\lambda_m~(m=1
\sim N)$ are eigenvalues. $N=n(n+3)/2$ for the
second order RG equations, while $N=n$ for the
first order RG equations, with
$n$ being the number of observables.   Since the maximum
(minimum) eigenvalue,
$\lambda_{\rm max}(\lambda_{\rm min})$ becomes
the most  relevant toward larger (smaller) scale,  the
scaling property of each observable around fixed point
is determined by
$\lambda_{\rm max}(\lambda_{\rm min})$.

The fixed points themselves are independent of the
scaling  parameter $s$, but the matrix  ${\cal
W}^{[2]}$ (or ${\cal
W}^{[1]}$) and its eigenvalues  depend on $s$.
For the sake of explicit argument below, we
restrict the range of the parameter
$s$ as
$0<s<1$ where the topology of the RG flow is not
changed \cite{foot4}.

\subsection{Fixed points and
the flow  of RG  equations without  fluctuations}
Here we  analyze the first order RG equations
(\ref{rerho})
$\sim$ (\ref{retidn-}) in detail.
First, by neglecting both  shear and
tidal force terms,  we have simple RG equations
(\ref{1st-RGeq}) with
\beq
A[\la F\ra ] =
\left(\begin{array}{c}
 -\la \theta\ra +
{\frac{1}{3}}\la
\theta\ra^2  + 4\pi G\la \rho\ra
\\[.5em]
(\la \theta\ra -2)
\la \rho\ra
\end{array}
\right)
\label{rerho2}
\eeq
for two averaged variables $\la F\ra =(\la
\rho\ra ,  \la \theta\ra)$.  We easily find three
fixed points
$\la F\ra^*_{(k)}$ ($k=1, 2, 3$), which are
listed  in Table 1.

The stability at each fixed point
$\la F\ra^*_{(k)}$ is examined by linearizing
the RG  equations around the fixed point (Eq.
(\ref{1st_perturb})).  All  eigenvalues of
${\cal W}^{[1]}$ for  three fixed points are
also given in Table 1.    For the first fixed
point (1) where both
$\la\theta\ra^*_{(1)}$ and $\la\rho\ra^*_{(1)}$
vanish, the  dynamical term
${\mit \Delta}_T \la F\ra$ does not affect the
scaling, so  that the eigen vector points  the
direction of each  observable.  On the other hand,
for other fixed  points,  the effect of dynamics
changes a scaling property  from the original bare
scaling  law  (\ref{fluscal}).
\par
As seen from Table 1, the fixed point (1) is stable
toward larger scale.
For increasing scale, the Universe approaches this
fixed point, at which  both the energy density and
expansion vanish. Therefore we shall
call it the fixed
point Q as the Universe is quiescent. The second
fixed point (2) is a saddle point. It represents the
``Einstein-de Sitter'' Universe because
$\la \theta\ra^*_{(2)} =
\theta_{\rm EdS}$ and $
\la \rho\ra^*_{(2)} = \rho_{\rm EdS}$. We then call it
the  fixed point E. The third fixed point (3) is
unstable in any direction. At first sight, it might
look strange since the expansion of the Universe is
finite  in spite of vanishing energy density. This
corresponds to the ``Milne" Universe in which the
expansion rate is apparently finite because of a
specific  choice of the time slice. We call it the
fixed point M.  This Universe is reduced to  the
``Minkowski'' space by adjusting the
time slice appropriately.
\par
 In Fig.1, to see more precisely those fixed points
and the behaviors of global structure of the RG
equations,  we depict the RG flow on the
$H$-$\Omega$ plane, where the
Hubble parameter
$H$ and the density parameter $\Omega$ are  defined
by
\barr
H&\equiv&{\la \theta\ra \over 3}
\\
\Omega&\equiv &{8\pi G\la\rho\ra \over 3H^2}.
\earr
The RG flow either approaches fixed point Q or
escapes to infinity  toward larger scale. These two
kinds of flows are clearly separated by the divide
which passes through two fixed points E and M
(Fig.1). The flow escaping to infinity seems
unphysical since both  $H$ and $\Omega$ blow up to
infinity at large scale.
%
%
\par
Next we include shear and tidal force, which
leads to  Eqs.(\ref{rerho}) $\sim$
(\ref{retidn-}).
The three fixed points Q, M and E found in the case
without shear and tidal force survive, and their
stabilities  also remain  unchanged (see Table 2).
In this sense,  the three fixed points  Q, M and E are
robust.  Four new  fixed points appear in
addition to  those three  fixed points, which
are listed in Table 2. Since all   new fixed
points are saddle points,  any RG flow  still
approaches the fixed point Q or escapes to
infinity toward larger  scale, or it approaches
the fixed point M
 toward smaller scale.
\subsection{Fixed points and the flow and  of RG
equations with  fluctuations}
%
%
%
Since our model of the Universe may be quite
inhomogeneous and could be fractal in a microscopic
point of view, an inclusion of fluctuations of the
physical variables is inevitably important, when we
compare our results with those for the real Universe.
In fact, if we do not
include fluctuations, it is mathematically equivalent to
a self-similar solution on a fixed time
slice \cite{foot5}, although
the physical meaning is quite different. Then, we
shall add fluctuations  and see  how the fluctuations
affect the RG flow  diagram. We just include   second
order cumulants to the RG equations and   truncate this
hierarchy there by ignoring    all  intrinsic
fluctuations (cumulants) higher than the second order,
although we may have another choice of higher-order
cumulants \cite{foot6}.

There are 27 independent variables
in this case and it becomes more difficult to find a
full list of  fixed points.
%
%
Then, first we reduce the number of variables and
start our  study  by the shear-free  case.  Since
independent observables are just
 $\rho$ and $\theta$ in this case,  the total number
of  variables including second order cumulants is
just five:
$\la \rho\ra $, $\la \theta\ra $,
$\la \rho^2\ra_{\rm c}$, $\la\theta^2\ra_{\rm c} $,
$\la\rho\theta \ra_{\rm c} $. We  list up all fixed
points  and  show their  stabilities in Table 3.
The  three important fixed points (1)$\sim$(3) in the
first order RG equations  still  remain and  their
stabilities are also unchanged. They are quite robust.
Since all of the new fixed points (8)$\sim$(10) are
saddle points,  global properties of RG flows toward
both larger and smaller scales are solely determined by
the fixed points  Q and M, respectively.  The fixed
point E   remains  as a saddle point.

As we mentioned, the fixed points
found in the first order equations are also those
in the second order with zero cumulants.  For the eigen
values at those fixed points, we find eigen values in
the first order are also those in the second order.
To see more detail, we  explicitly write down the
perturbation matrix ${\cal W}^{[2]}$
(\ref{2nd-matrix}) as\\[1em]
\[
 {\cal W}^{[1]} \equiv \left( \begin{array}{cc}
 w_{11}  & w_{12}
\\[.5em]
  w_{21} &
 w_{22}
\end{array}
\right)
= {\cal S}^*\left( \begin{array}{cc}
 \langle \theta
\rangle^* -2  & \langle\rho\rangle^*
\\[.5em]
  4\pi G &
 {\frac{2}{3}} \langle \theta \rangle^* -1
\end{array}
\right) , ~~~~~~~~{\cal S}^* C = {\cal S}^* \left(
\begin{array}{ccc} 0&0&0
\\[.5em]
0&-{2\over 3}&0
\end{array}
\right)
\]
\vskip .5cm
\beq
 {\cal S}^*B[\la F\ra^*] =
{\cal S}^*\left( \begin{array}{ccc}
2(\langle \theta
\rangle^*-2) &0&2\langle \rho \rangle^*\\[.2em]
0&2\left({2
\over 3}\langle
\theta
\rangle^*-1\right)&8\pi G\\[.2em]
4\pi G&\langle\rho \rangle^*&
{5 \over 3}\langle \theta
\rangle^*-3
\end{array}
\right)  =
\left( \begin{array}{ccc}
2w_{11} &0&2w_{12}\\[.2em]
0&2w_{22}&2w_{21}\\[.2em]
w_{21}&w_{12}&
w_{11}+w_{22}
\end{array}
\right)
\eeq
\vskip .5cm
\noindent
Then, the eigenvalue equation for the second order
perturbation matrix ${\cal W}^{[2]}$ is
\beq
(\lambda^2 - {\rm tr} {\cal W}^{[1]}
 \cdot
\lambda + {\rm det}  {\cal W}^{[1]})
(\lambda- {\rm tr}{\cal W}^{[1]}
) (\lambda^2 - 2~{\rm tr}{\cal W}^{[1]}
\cdot
\lambda + 4~{\rm det} {\cal W}^{[1]}
 )= 0.
\label{2nd-eigen}
\eeq
Since the eigenvalue equation for the first order
matrix ${\cal W}^{[1]}$ is
\beq
 \lambda^2 - {\rm tr} {\cal W}^{[1]}
 \cdot
\lambda + {\rm det}  {\cal W}^{[1]}
 = 0,
\label{1st-eigen}
\eeq
supposing that  $\lambda^{[1]}_1$ and $\lambda^{[1]}_2$
are its two solutions, we find the
solutions of Eq. (\ref{2nd-eigen}),
$\lambda^{[2]}_i ~(i=1\sim 5)$, as
\barr
\lambda^{[2]}_1 &=& \lambda^{[1]}_1, ~~~~
\lambda^{[2]}_2 ~=~ \lambda^{[1]}_2, \nonumber \\
\lambda^{[2]}_3 &=& 2 \lambda^{[1]}_1, ~~~~
\lambda^{[2]}_4 ~=~ {\rm tr}  {\cal W}^{[1]}
 = \lambda^{[1]}_1+\lambda^{[1]}_2,  ~~~~
\lambda^{[2]}_5 ~=~ 2\lambda^{[1]}_2 .
\label{1st-2nd_1}
\earr
Therefore the minimum eigenvalues at fixed
points in the first order, in particular the fixed
points Q, E and M, are given by
\beq
\lambda^{[2]}_{\rm min} = 2\lambda^{[1]}_{\rm min} ,
\eeq
if $\lambda^{[1]}_{\rm min}<0 $ (this is the case for the
fixed points Q and E), otherwise
\beq
\lambda^{[2]}_{\rm min} = \lambda^{[1]}_{\rm min} ,
\eeq
(this is the case for the fixed point M).
%
%
\par
Finally, we study a full model with 27 RG
equations for 6 independent variables
including shear and tidal force and their second
order cumulants.  Though
it is very  difficult to find all fixed  points,
all seven fixed points in the first order RG
equations  (listed in Table 2) remain as fixed
points, as we mentioned. As for their eigenvalues,
we also find the same result as that in the case
without shear and tidal force, that is, the
eigenvalue in the second order RG equations
\{$\lambda_{m}^{[2]} | m=1,
\cdots , n(n+3)/2$\} are
constructed by those in the first
order $\{\lambda_m^{[1]} | m=1, \cdots , n$ \} as
\barr
\lambda_{m}^{[2]} & = &
\lambda_{m}^{[1]} ~~~~{\rm for }~~ m=1,
\cdots , n \nonumber  \\
\lambda_{m}^{[2]} & = &
\lambda_{l}^{[1]}+\lambda_{l'}^{[1]}
~~~{\rm for } ~~ m=n+1,\cdots , n(n+3)/2 ~~{\rm
with}~~ l, l' = 1, \cdots , n
\label{1st-2nd_2}
\label{lambda2}
\earr
where $n$ is the number of observables.
This result is understood from some special
relation between the perturbation matrices ${\cal
W}^{[1]}$ and
${\cal S}^*B[\la F\ra^*]$ \cite{foot7};
\begin{eqnarray}
{\cal
S}^*B_{ij,kl}[\la F\ra^*] ={1 \over 2} \left[
\delta_{ik}w_{jl}
+\delta_{il}w_{jk}
+\delta_{jk}w_{il}
+\delta_{jl}w_{ik} \right],
\end{eqnarray}
where $i,j,k,l$ denote the observables $\rho,
\theta,
\sigma_\pm, E_\pm$.

In Table 4, we have only listed three important
fixed  points Q, E and M with  their stabilities.
These three fixed points still survive and their
stabilities remain unchanged
 even after including  fluctuations as well as shear
and tidal force.
%
%
\par
Let us just mention how our results change if  we
use  another  approximation scheme
 for the equation of motion for $E_{ij}$.
We mainly use LTA approximation in this paper.
In the case of other local approximations
like NMA \cite{foot8}, 
the RG equations become more complicated and it
is more  difficult to look for all  fixed points
and eigenvalues.  However, it is still  easy to
show that the above three fixed points Q, M and E
survive since the RG equations exactly reduce to
the shear-free case  when we set
$E_\pm =\sigma_\pm =0$.  The stabilities of these
fixed  points can also be checked by examining
the RG flow around them.  We  show that the
stabilities of these fixed points remain
unchanged  even in the NMA. The minimum
eigenvalue, which is the most relevant toward
smaller scale, is also invariant irrespective of
local approximation scheme of $E_{ij}$, because
 the eigen-direction for $\lambda_{\rm
min}^{[2]}$ (or $\lambda_{\rm min}^{[1]}$) is
orthogonal to both shear and tidal directions in
27-dimensional ``phase" space of $(\la F\ra , \la
FF\ra_{\rm c})$ (or in 6-dimensional ``phase" space
of
$\la F\ra$). Hence the  effect of shear or tidal
force seems irrelevant to the scaling property of
observables around fixed point E or M toward
smaller scale.
\par
 The  three fixed points E, M, and Q
are robust and  change neither their stability
nor  the scaling property of observables around
them toward smaller scale, regardless of shear,
tidal force, and then the approximation scheme of
the tidal force.
\subsection{The flow of our Universe}
%
%
The RG flow and the fixed points we argued in the
previous subsections describe all the possible
diversity of our system.  Only a part of them, i.e. a
particular integral line in the flow  diagram  which
satisfies certain boundary condition,   actually
describes our real Universe. In order to elucidate
such integral flow, we need a  plausible
boundary condition and examine the global/local
scaling properties of observables.
\par
One plausible boundary condition at  the present horizon
scale and beyond could be the fixed  point E
(``Einstein-de Sitter" Universe).   This boundary
condition would be inferred from the inflationary
scenario in the early Universe, which
 predicts   the flat FRW Universe  with tiny
fluctuations as our real Universe.

With this
boundary condition, we solve the RG equations  toward
smaller scale and study the scale change of various
observables.   In solving the RG equations  toward  the
smaller scale, we must remember that (i) the positive
(negative) eigenvalue of the  matrix
${\cal W}^{[2]}$ at a fixed point denotes stability
(instability). (ii) The direction of the integration is
opposite to that of flow indicated in Fig. 1. It is clear
that  the choice of the exact values of point E as a
boundary condition is trivial and not interesting because
 it is the fixed point  and the flow stays there
forever.  To find more interesting model of the
Universe,   we need to start from the close vicinity of
the fixed point E.

 Here we choose the boundary condition as follows.
For the averaged variables, we use the exact
value of the fixed point E:
\barr
\la \theta\ra = \theta_{\rm EdS} =2, ~~
\la \rho\ra = \rho_{\rm EdS}=1/6\pi G ,
\la \sigma_\pm\ra =\la E_\pm\ra =0,
\earr
which gives $\Omega=1$. Since we expect some
fluctuations even in an inflationary model, we set small
finite fluctuations at  the  fixed E point.  We assume
that the off-diagonal fluctuations have  initially no
intrinsic fluctuations \cite{foot9}, i.e.
\barr
\la f_if_j\ra_{\rm c}=0 \ (i\neq j)
~~({\rm at}~~\ell =0).
\earr

Before going to show our numerical result,
we shall discuss the asymptotic behaviors near three
robust fixed points, Q, E, and M.
We showed that the scaling property
of observables around those fixed  points toward smaller
scale is determined by the  minimum eigenvalue
$\lambda^{[2]}_{\rm min}$.
This is always true for the second order cumulants $\la
f_if_j\ra_{\rm c}$. The scaling
property of the averaged variable
$\la f_i\ra$ may, however,  depend on the case,
because the eigen vector with the eigenvalue
$\lambda^{[1]}_{\rm min}$ is perpendicular to the
hypersurface of the cumulants
in the ``phase" space $(\la F \ra , \la FF\ra_{\rm c})$.
If we have fluctuations  initially rather
than a deviation from the fixed point, i.e.
  $\la f_i^2\ra^*_{\rm c} \neq 0$, then
\beq
\la f_i\ra \sim \la f_i\ra^* +
\epsilon_i e^{\lambda^{[2]}_{\rm min}\ell},
\label{asymptotic}
\eeq
as $\ell$ decreases, unless
$\la f_i^2\ra^*_{\rm c}
\gsim (\la f_i\ra^*)^2$. This is just because how to
leave from the fixed point is determined by the fluctuations
$\la f_i^2\ra^*_{\rm c} $, which scaling property is fixed
by
$\lambda^{[2]}_{\rm min}$.
 On the
other hand,   if we have an initial
deviation from the fixed point as $\la f_i\ra \neq \la
f_i\ra^*$ without the initial cumulant ($\la
f_i^2\ra^*_{\rm c} =0$), then
$\la f_i\ra - \la f_i\ra^* $ behaves  first as $
\exp(\lambda^{[1]}_{\rm min}\ell)
$.  Since our boundary condition is the former case, we
expect the asymptotic behavior as Eq. (\ref{asymptotic}).

It may be more comprehensive to show our results
if we introduce the
``dispersions" $\Delta_{ij}$  instead of the cumulants
$\la f_if_j\ra_{\rm c}$.
For the diagonal cumulants $\la f_i^2\ra_{\rm c}$,  we
define them as
\beq
\Delta^2_{\rho} = \la \rho^2\ra_{\rm c} / \la
\rho\ra ^2 ,
~
\Delta^2_{\theta} = \la\theta^2\ra_{\rm c} / \la
\theta\ra ^2 ,
~
\bar{\Delta}^2_{\sigma_\pm} =
\la \sigma_\pm^2\ra_{\rm c} / \la
\theta\ra ^2 ,
~
\bar{\Delta}^2_{E_\pm} =  \la E_\pm^2\ra_{\rm c} /\la
\theta\ra^4
\label{disper2},
\eeq
where for shear and tidal
force, $\bar{\Delta}_{\sigma_\pm}$ and
$\bar{\Delta}_{E_\pm}$, which are normalized by $\la
\theta\ra$, have been
used rather than the conventional dispersions because
those averaged variables sometimes vanish even if their
cumulants are finite, resulting in a
divergence of the dispersions.  The dispersion parameters
$\Delta_{\rho},~ \Delta_{\theta},~
\bar{\Delta}_{\sigma\pm},~
\bar{\Delta}_{E\pm}$ are assumed to be sufficiently small
at the boundary $\ell =0$.
In what follows, just for simplicity, we also set
$\Delta_{\rho}=
\Delta_{\theta}\equiv\Delta_0$ at $\ell =0$, which
may be plausible  because
$\rho$  and
$\theta$ may be closely connected with each other
through the  fluid equations and are expected to have
the same order of fluctuations.

\subsubsection{The case without shear and tidal force}
First we consider the case without shear and tidal
force. In
Fig.2, we depict the scale dependence of $\Omega , H,
\Delta_\rho , \Delta_\theta$ in terms of $L\equiv
e^\ell $, which is the ratio of the averaging scale to the
horizon scale (or the scale at our boundary). Since we
have chosen $\ell =0$ as the  horizon scale (or beyond),
our observable Universe is described by  negative
$\ell $.  As seen from Fig.2, the
integrated flow line  always  converges to the values at
fixed point M, i.e.
$\Omega_{\rm Milne}=0$ and $H_{\rm Milne}=1$.  This
means that the ``Milne" Universe is the unique stable
 fixed point toward smaller scale. Once it comes close to
fixed  point M, the behavior of the RG flow at small
scale  is determined by the minimum eigenvalue at
M, i.e. $\lambda^{[2]}_{\rm min}=s/(1-s)$.

To see a role of the scaling parameter $s$, in Fig.3, we
plot the scale dependence of the density parameter
$\Omega$, the expansion $H$, and the fluctuations
$\Delta_\rho, \Delta_\theta$
 for several values of $s$ with fixed initial
fluctuations
$\Delta_0=10^{-2}$.
We find that the observables  show the
following universal behavior: \\
(i) $H$ increases  toward smaller scale from $2/3$
to $1$.  \\
(ii) $\Omega$
reduces from $1$ to $0$ toward smaller scale. There is a
plateau near the fixed point E, while a constant slope
is found near the fixed point M. 
\par
The asymptotic values are given by those at the fixed
points E and M.
We can also explain a  constant slope at the
smaller scale and   a plateau at larger scale in
$\Omega$ diagram.
Since $\la\rho\ra \sim O( \exp \{[s/(1-s)] \ell
\})$ and $\la\theta \ra \sim 3$  near the fixed
point M  $\Omega$ at smaller scale behaves as
\beq
\Omega \sim O\left(
\exp\{ [s/(1-s)] \ell\}\right)  ,
\eeq
where $s/(1-s)$ is the minimum eigenvalue
evaluated at the fixed point M.
Then   the
slope of the $\Omega$-curve near  M
turns out to be
\beq
{d \log \Omega \over d \log L} \sim {s \over
1-s}.
\eeq
On the other hand, the plateau appears near the fixed
point E, because
\beq
\la \rho\ra /\rho_{\rm EdS} \sim 1  + \epsilon_\rho
 \exp \{ -[2s/(3-2s)]  \ell \},  ~~~~
\la\theta\ra /\theta_{\rm EdS}\sim 1
+ \epsilon_\theta
 \exp \{ -[2s/(3-2s)]   \ell \}  \eeq
 and then
\beq
\Omega \sim 1
+ (\epsilon_\rho - 2\epsilon_\theta )
 \exp [-(2s/(3-2s))  \ell ]  \approx 1,
\eeq
where we
have used
$\lambda^{[2]}_{\rm min} = -2s/(3-2s)$ evaluated at E.
The turning point $L_{\rm cr}$
of the slope into the plateau is
estimated easily.  Since the deviations from the EdS
values are due to the fluctuations  $\Delta_0$, $
\epsilon_\rho ,  \epsilon_\theta  \sim O(\Delta_0)$.
Then, when
$\Delta_0
\exp \{-[2s/(3-2s)]
\ell_{\rm cr}  \} \sim O(1)$, i.e.
\beq
\ln L_{\rm cr} = \ell_{\rm cr} \sim [(3-2s)/2s]\ln
\Delta_0 ,
\eeq
 we find a large deviation from the EdS
values.    Thus the minima of the eigen values at two
fixed points E and M determine the asymptotic structure
of our Universe.

Is there any observational data which may suggest a
scale dependence of either $\Omega$ or density? In
discussion of dark matter, when we plot the
mass-luminosity ratio in terms of the scale of the
objects, we find a similar curve to Fig. 3(a). 
If we
assume that the mass-luminosity ratio curve is what we
are discussing here, we find some constraint of $s$ as
$0.3 ~\lsim ~s ~\lsim 0.7$.

\subsubsection{The case with shear and tidal force}
Next, we examine the effect of shear and tidal force.
Just for simplicity, we assume that
$\sigma_-=E_-=0$, which is consistent with the
RG equations.  As we assumed, $\la\sigma_+\ra$ and $\la
E_+\ra$ vanish at $\ell =0$. Then initial free
parameters are $\Delta_0, \bar{\Delta}_{\sigma_-, 0}$ and
$\bar{\Delta}_{E_-, 0}$, which are fluctuations at $\ell
=0$. Our numerical  calculations, shown in Figs. 4, 5, 
reveal that they tend to prevent
$H$ and $\Omega$ from converging to the fixed point  M
 though ``Milne" Universe is still a stable
fixed point. More precisely,  we observe three
qualitatively different behaviors of the RG flow  of our
Universe:
\begin{itemize}
\item[(i)]
The RG flow monotonically converges to fixed point M.
$H$ and $\Omega$ approach to the value determined by
fixed  point M toward smaller  scale.
\item [(ii)]
Initially the Universe comes near the fixed M point,
resulting that
$H$  increases and $\Omega$  decreases toward smaller
scale. However it turns around before reaching the fixed
point M   and eventually diverges to infinity in the
flow diagram as $H \rightarrow -\infty$ and $\Omega
\rightarrow \infty$.
\item [(iii)]
The RG flow monotonically diverges to infinity in the
flow  diagram as $H
\rightarrow -\infty$ and $\Omega \rightarrow \infty$.
\end{itemize}
\par
We can classify the parameter space
$(\Delta_0,\,\bar{\Delta}_{\sigma +,0},\,\bar{\Delta}_{E
+,0})$ of   fluctuations at the boundary (near the
horizon) into three regions according  to the above
asymptotic behaviors of the RG flow. The class (i) is
realized if the  fluctuations of shear and tidal force are
sufficiently smaller than those of the density and the
expansion, i.e.
$\bar{\Delta}_{\sigma_+,0}/\Delta_0
\sim \bar{\Delta}_{E_+,0}/\Delta_0<R_{cr}$.  Otherwise,
i.e. if
$\bar{\Delta}_{\sigma_+,0}/\Delta_0
\sim \bar{\Delta}_{E_+,0}/\Delta_0 > R_{cr}$,
 the class (iii) will happen. The critical value
$R_{cr}$ is almost independent of
$\Delta_0$ and approximately 
$R_{cr}\approx 0.15$. The class (ii)  is a
narrow boundary  between them. Hence, the class (i) (or
(ii)) naturally occurs as long as the shear  and tidal
force are sufficiently smaller than the density
fluctuations at the horizon  scale. This situation is
strongly suggested by recent observations
\cite{kogut97}, i.e. the  shear is strictly restricted
around
$\sigma/H < 10^{-9}$ by CMB data.  Hence we expect that
our Universe belongs to the class (i) and its flow line
converges into the fixed point M in smaller scale.
The $s$ dependence of $\Omega$ or $H$ is quite similar to
the case without shear and tidal force (Fig. 6 (a), (b)).

As for the shear and tidal force, their averaged values
increase toward smaller scale, but eventually decrease if
the flow will converge to the fixed point M (Fig. 6 (c), (d)).
 The scaling properties of   the shear $\la
\sigma_+\ra $ and  the tidal force $\la  E_+\ra $  are
also  determined by
$\lambda^{[2]}_{\rm min}$ at the  fixed points E and M.
 Both shear and tidal force are dumped off at
smaller scale along the flow converging to the fixed
point M.

The behaviors of
fluctuations
$\Delta_\rho$, $\Delta_\theta$,
 $\bar{\Delta}_{\sigma_+}$ and
$\bar{\Delta}_{E_+}$   are shown in Fig. 6 (e)$\sim$(h).
We find that the similar behaviors to the case without
shear and tidal force for $\Delta_\rho$
and $\Delta_\theta$.

\subsection{Beyond the horizon scale }
In the previous subsections, we have integrated  the RG
equations from the horizon scale ($\ell$=0)  toward
the smaller scale.  It is also possible to solve them
toward the larger scale with the same boundary
conditions. The structure beyond the horizon scale is
not observable at present  but may give some relation
with an inflationary scenario in the early Universe.
In Fig. 7, we plot the scale dependence of
$\la\rho\ra$, $H$ and those 
fluctuations $\Delta_{\rho}$ and $\Delta_{\theta}$
in shear-free case with initial fluctuation
$\Delta_0=10^{-2.5}$. In the finite scale $10^{-5}<L<10^5$,
the flow stays in the vicinity of fixed point E and
the power law behaviors of $\Delta_{\rho}$ and $\Delta_{\theta}$
toward larger scale
are determined by $\lambda^{[2]}_{\rm max}$ at fixed point E,
while in larger scale $L>10^5$, the flow converges to
fixed point Q after $H$ becomes
negative and their power law behaviors  are
characterized by the eigenvalues  of ${\cal
W}^{[2]}$ at the fixed  point Q. These eigenvalues
are the same ones naively determined by
the original  scaling. 
This is consistent with the
fact that the expansion  ceases at the  fixed
point Q and the RG transformation reduces to a
pure scale transformation. This property makes
$\Delta_{\rho}$ constant around
fixed point Q because of the cancellation
between numerator and denominator in the definition
of (\ref{disper2})
 (Fig. 7 (c)), while
$\Delta_{\theta}$ diverges at the scale where $H$ 
vanishes and turns
negative from the definition (\ref{disper2}) (Fig. 7 (d)).
 The declination of the
slope in the
$\log\la \rho\ra$-$\log L$ diagram (Fig. 7 (a))  is
$-2s$ which is the minimum eigenvalue of ${\cal
W}^{[2]}$ associated with the fixed point Q.  If
we assign a fractal dimension
$D$ for this Universe, $D=3-2s$ because $\la \rho\ra$
reduces as $e^{(D-3)\ell}$ from a fractal structure
while $\la \rho\ra
\propto e^{-2s\ell }$ by our scaling law. If we
assign the value
$s=0.5$, the
corresponding fractal dimension becomes $D=2$.
Because our model is a Newtonian  cosmology, further
analysis beyond the horizon scale would be irrelevant
to  the real Universe.
%
\section{Concluding Remarks}
\par
The renormalization group method has been applied to
Newtonian cosmology  in this paper and the scale
dependence of the averaged observables on a fixed
time slice  including effects of fluctuations has been
studied.  The scaling assumption we proposed for the
averaged observables is fulfilled in the case of
either homogeneous distribution or non-analytic
distribution such as a fractal, if we assume such a
scaling at any  position and at fixed time.  Actual
observation of the two point correlation function of
galaxies may favor such a fractality at least  in the
finite scale range below 10 Mpc. Multi-fractality
could be applied to explain  the transition from
finite fractal range at smaller scale  to  homogeneous
distribution at larger scale\cite{Sch89}.
 Our scaling
analysis of averaged observables here might give a
hint to the behavior of observables in such a fractal
matter distribution, if it exists.

To close our dynamical system and apply the RG method,
here we have adopted an approximation of tidal force 
(LTA) and considered only fluctuations up to
the second order. The higher order cumulants of
averaged observables are ignored.
 We have found three robust fixed points of  the RG
equations.\\
(1) The fixed point Q (``Quiescent" Universe) is
stable  in any  directions toward larger scale.\\
(2) The fixed point E (``Einstein-de Sitter" Universe)
is a saddle point.\\
(3) The fixed point M (``Milne" Universe) is
unstable in any  directions toward  larger scale.

These fixed points do not change their stability even if we
extend  the ``phase" space including fluctuations as
well as shear and
tidal force.  Any other fixed points are saddle, and
therefore the scaling property of the global RG flow
toward larger  (smaller) scale is determined by  fixed
point Q (M).  We find that the Universe asymptotically
approaches fixed point Q or diverges toward
infinitely regardless of the detail at the smaller
scale.

In order to find out the flow line which
represents  our real Universe, we have imposed the
boundary condition fixed point E with tiny
fluctuations at the horizon scale. The inflationary
scenario in the early Universe would imply such a
boundary condition, i.e. the inflation predicts the
Einstein-de Sitter Universe at the horizon scale and
beyond. We have solved the RG equation toward smaller
scale from this boundary. Then we find  that, toward
smaller scale, the expansion $H$  defined by
$\la \theta \ra /3$ monotonically increases from $2/3$ to
$1$, and the density parameter $\Omega = 8\pi G
\la\rho\ra /3H^2$ monotonically reduces from $1$ to
$0$,  provided the
fluctuations of shear and tidal force are much
smaller than the density fluctuations, which would be
supported by the inflationary scenario
\cite{barrow97} and observations of CMB.

\par
We have two important free parameters, $s$ and $\Delta_0$.  These
values might be fixed by some of the following
arguments:\\
(i) The scale dependence of the
Mass-Luminosity ratio of astrophysical objects.\\
(ii) A systematic
decrease of the Hubble parameter $H$, if any, toward
the larger scale.\\
(iii) The scale dependence of the amplitude of the
two point correlation functions.\\
(iv) Some of the
inflationary models  predict a fractal-like structure
beyond the horizon scale
\cite{vilenkin84}.
\par
Several comments related with our future work are  in
order.\\
(a) We definitely need the general relativistic
generalization of  our work.  This is because the
large scale structures at high redshift and beyond
the horizon scale would be adequately described only
by general relativity.\\
(b) In our analysis, we assume a scaling property
of the averaged observables and their fluctuations.
However we do not know whether such a solution is
dynamically stable or not. To answer for this, we have to
consider the dynamical stability of our system.\\
(c) So far we have considered fluctuations
of variables up to the second order. 
Though this 
would be sufficient when we consider the vicinity of the 
horizon scale,  we ultimately need to incorporate
higher order fluctuations  as well for the consistency
of the theory.  
\\  
(d) We have used the Local Tidal Approximation (LTA) for obtaining equations of motions for tidal force.
By this method, we could avoid directly solving Poisson equation.  
This approximation would be justified upto mildly non-linear regime.
However it may not be valid in the fully non-linear regime, 
especially in smaller scales, 
where the fluctuations grow $\Delta_{\rho} \gg 1$.
In order to treat the smaller scales properly, we need to improve LTA in our future study.    
\\
 We would like to report the analysis on these
issues in the near future.

\vskip 1.5cm
\noindent
We would like to thank Paul Haines, Hiraku Mutoh, Kamilla
Piotrkowska, Naoshi Sugiyama, Takayuki Tatekawa, and Kenji
Tomita  for many useful discussions.
This work was supported partially by the Grant-in-Aid
for Scientific Research Fund of the Ministry of
Education,  Science and Culture (Specially Promoted
Research No.  08102010),  and by the Waseda University
Grant for Special Research Projects.
%
%

\oddsidemargin -0.3in
\doublerulesep=0pt

\newpage
~\\
\vskip 2cm
\begin{center}
\def\arraystretch{2.0}
\begin{tabular}{| l || c c | c | c |}
\hline\hline\hline &
$6\pi G\la\rho\ra^*$& $\la\theta\ra^*$& ${\cal S}^*$
& two eigenvalues \\
\hline\hline
~(1) Q ~& ~0 & 0 &$s$& \bigg\{$-2{\cal S}^*,-{\cal
S}^*$\bigg\}
\\
\hline ~(2) E ~& ~1 & 2 &$\displaystyle\frac{3s}{3-2s}$&
$\bigg\{
\displaystyle -\frac{2}{3}{\cal S}^*,
{\cal S}^*\bigg\}$ \\   \hline
~(3) M~ & ~0 & 3 &$\displaystyle\frac{s}{1-s}$&
$\bigg\{ \displaystyle {\cal S}^*, {\cal S}^*
\bigg\}$
\\ \hline
\end{tabular}
\end{center}
\vskip .5cm

\begin{flushleft}
Table 1. The list of three fixed points and their
eigenvalues  in the case  with neither shear and tidal
force nor  fluctuations.
\\
\end{flushleft}

\vskip 2cm
\begin{flushleft}
\def\arraystretch{2.0}
\tabcolsep=4pt
\begin{tabular}{| l || c c c c c c | c | c |}
\hline\hline\hline
 &
$6\pi G\la\rho\ra^*$ & $~~\la\theta\ra^* ~$&
$\la\sigma_+\ra^*$ & $\la\sigma_-\ra^*$ &
$\la E_+\ra^*$ & $\la E_-\ra^*$ &${\cal S}^*$&
six eigenvalues
\\
\hline\hline\hline
~(1) Q~ & ~0 & 0 & 0 & 0 & 0 & 0 &$\displaystyle s$&
\bigg\{$-2{\cal S}^*,-2{\cal S}^*,-2{\cal S}^*,-{\cal
S}^*,-{\cal S}^*,-{\cal S}^*$\bigg\} \\
\hline ~(2) E~
&~1&2&0&0&0&0&$\displaystyle\frac{3s}{3-2s}$&
$\bigg\{\displaystyle -\frac{2}{3}{\cal S}^*,
-\frac{2}{3}{\cal S}^*,
-\frac{2}{3}{\cal S}^*, {\cal S}^*,
{\cal S}^*, {\cal S}^*\bigg\}$ \\ \hline
~(3) M~
&~0&3&0&0&0&0&$\displaystyle\frac{s}{1-s}$&
$\bigg\{\displaystyle {\cal S}^*,
{\cal S}^*,{\cal S}^*,
{\cal S}^*,{\cal S}^*,{\cal S}^*\bigg\}$ \\
\hline
~(4)&~0&1&$\displaystyle
-\frac{1}{3}$&$0$&0&0&$\displaystyle\frac{3s}{3-s}$&
$\bigg\{\displaystyle -{\cal S}^*, -{\cal S}^*,
-{\cal S}^*, -{\cal S}^*,
-{\cal S}^*, {\cal S}^*\bigg\}$ \\  \hline
~(5)&~0&1&$\displaystyle \frac{1}{6}$&
$\displaystyle
\frac{1}{2\sqrt{3}}$&0&0&$\displaystyle\frac{3s}{3-s}$&
$\bigg\{\displaystyle -{\cal S}^*, -{\cal S}^*,
-{\cal S}^*, -{\cal S}^*,
-{\cal S}^*, {\cal S}^*\bigg\}$ \\ \hline
~(6)&~0&1&$\displaystyle \frac{1}{6}$&
$\displaystyle
-\frac{1}{2\sqrt{3}}$&0&0&$\displaystyle\frac{3s}{3-s}$&
$\bigg\{\displaystyle -{\cal S}^*, -{\cal S}^*,
-{\cal S}^*, -{\cal S}^*,
-{\cal S}^*, {\cal S}^*\bigg\}$ \\ \hline
~(7)&~0&2&$\sigma_+^*$&$\sigma_-^*$&
$E_+^*$ &$E_-^*$&$\displaystyle\frac{3s}{3-2s}$&
$\bigg\{\displaystyle
\bar{\lambda}_1  {\cal S}^*,
\bar{\lambda}_2  {\cal S}^*, 0, 0, {\cal S}^*,
 \bar{\lambda}_3  {\cal S}^*\bigg\}$ \\
\hline
\end{tabular}
\end{flushleft}
\vskip .5cm

\begin{flushleft}
Table 2. The list of 7 fixed points and their eigenvalues
in the case without fluctuations
in LTA.  The seventh fixed point [7] is not an isolated
point  but a closed loop parameterized by
one parameter $\eta$ as $  \sigma_+^*=
\frac{1}{3}\cos\eta$, $
\sigma_-^*=\frac{1}{3}\sin\eta$, $ E_+^*=\frac{1}{9}
(\cos 2\eta -3 \cos \eta )$, $
 E_-^*=-\frac{1}{9} (\sin 2 \eta +3 \sin \eta)
 $, where $0\leq \eta <2\pi$. $\bar{\lambda}_1 ,
\bar{\lambda}_2 , \bar{\lambda}_3$   are
three real  solutions $ (-1\leq
\bar{\lambda}_1 <
\bar{\lambda}_2\leq 0, \bar{\lambda}_3 \geq 1)$ of
$\bar{\lambda}^3-\bar{\lambda}
 - 4(1-\cos 3\eta
)/27 = 0$.
\end{flushleft}

\newpage
~\\
\vskip 2cm
\begin{flushleft}
\def\arraystretch{2.0}
\tabcolsep=4pt
\begin{tabular}{| l || c c c c c | c | c |}
\hline\hline\hline  & $6\pi G\la\rho\ra^*$ &
$\la\theta\ra^*$&
$36\pi^2 G^2\la\rho^2\ra^*_{\rm c}$&
$\la\theta^2\ra^*_{\rm c}$ & $6\pi
G\la\rho\theta\ra^*_{\rm c}$  &${\cal S}^*$&
five eigenvalues
\\
\hline\hline\hline
~(1) Q~ & ~$0$ & $0$ & $0$ & $0$ & $0$ &$s$&
$\bigg\{ -4{\cal S}^*,-3{\cal S}^*,-2{\cal
S}^*,-2{\cal S}^*,-{\cal S}^*\bigg\}
  $  \\ \hline

~(2) E ~& ~$1$ & $2$ & $0$ &$0$ &
$0$ &$\displaystyle\frac{3 s}{3 - 2s}$&

$\bigg\{
{\displaystyle -\frac{4}{ 3}{\cal S}^*},
{\displaystyle -\frac{2}{ 3}{\cal S}^*},
{\displaystyle \frac{1}{ 3}{\cal S}^*},
{\displaystyle {\cal S}^*},
{\displaystyle 2{\cal S}^*}
\bigg\}$
\\ \hline

~(3) M~ & ~$0$ & $3$ & $0$ &$0$ & $0$
&$\displaystyle\frac{s}{1 - s}$&
$\bigg\{
{\displaystyle {\cal S}^*},
{\displaystyle {\cal S}^*},
{\displaystyle 2{\cal S}^*},
{\displaystyle 2{\cal S}^*},
{\displaystyle 2{\cal S}^*}
\bigg\}$
\\ \hline

~(8)  & ~$0$& ${\displaystyle\frac{3}{2}}$ &
$0$ &${\displaystyle
-\frac{9}{8}}$ & $0$ &$\displaystyle\frac{2 s}{2 - s}$&
$\bigg\{
-{\displaystyle {\cal S}^*},
-{\displaystyle {\cal S}^*},
 {\displaystyle -\frac{1}{2}{\cal S}^*},
{\displaystyle -\frac{1}{2}{\cal S}^*},
{\displaystyle {\cal S}^*}
\bigg\} $ \\
\hline

~(9)  & ~$0$& ${\displaystyle\frac{9}{5}}$ &
$0$ &${\displaystyle -\frac{27}{25}}$ &
${\displaystyle\frac{81}{250}}$
&$\displaystyle\frac{5 s}{5 - 3 s}$&
$\bigg\{
{\displaystyle  -\frac{1 + \sqrt{13}
}{5}{\cal S}^*},
{\displaystyle -\frac{2}{5}{\cal S}^*},
{\displaystyle -\frac{1}{5}{\cal S}^*},
{\displaystyle\frac{ -1 + {\sqrt{13}}
}{5}{\cal S}^*},
{\displaystyle {\cal S}^*}
\bigg\} $
\\ \hline

(10)& ~$0$  & $2$ & $-{\displaystyle
\frac{1}{4}}$ &$-1$ & ${\displaystyle\frac{1}{2}}$
&$\displaystyle\frac{3 s}{3 - 2s}$&

$\bigg\{{\displaystyle  -\frac{1 + {\sqrt{17}}
}{6}{\cal S}^*}, 0,{
 \displaystyle \frac{ -1 + {\sqrt{17}} }{6}{\cal S}^*},
{\displaystyle
\frac{2}{ 3}{\cal S}^*},{
   \displaystyle {\cal S}^*}\bigg\} $  \\
\hline
 \end{tabular}
\end{flushleft}
\vskip .5cm

\begin{flushleft}
Table 3. The list of fixed points and their eigenvalues
 in the case  without shear and tidal force including
fluctuations. New fixed points (8) $\sim$ (10) are
unphysical because  $\la\rho^2\ra_{\rm c}^*$ and/or
$\la\theta^2\ra_{\rm c}^*$ are negative.
 \end{flushleft}

\vskip 2cm
\begin{center}
\def\arraystretch{2.0}
\tabcolsep=2pt
\begin{tabular}{| l || c c c | c | c |}
\hline\hline\hline
 & $~6\pi G\la\rho\ra^*$ & $~~\la\theta\ra^*
~$&
$\la {\rm others}\ra^*$&${\cal S}^*$& 27 eigenvalues
\\ \hline\hline\hline
~(1) Q~ &~0&0&0&$s$& $\bigg\{ (-4{\cal S}^*)\times 6,
(-3{\cal S}^*)\times 9, (-2{\cal S}^*)\times 9,
(-{\cal S}^*)\times 3
\bigg\}$ \\
\hline
~(2)
E~ &~1&2&0&$\displaystyle \frac{3s}{3-2s}$& $\bigg\{
\left(\displaystyle -\frac{4}{3}{\cal S}^*\right)\times
6,
\left(\displaystyle
 -\frac{2}{3}{\cal S}^*\right)\times 3,
\left(\displaystyle
\frac{1}{3}{\cal S}^*\right)\times 9,
\left(\displaystyle {\cal S}^*\right)\times 3,
\left(\displaystyle 2{\cal S}^*\right)\times
6\bigg\}$
\\ \hline
~(3) M~ &~0&3&0&$\displaystyle \frac{s}{1-s}$& $\bigg\{
\left(\displaystyle {\cal S}^*\right)\times 6 ,
\left(\displaystyle
2{\cal S}^*\right)\times 21
\bigg\}$
\\ \hline
\end{tabular}
\end{center}
\vskip .5cm

\begin{flushleft}
Table 4. Three important fixed points and their
eigenvalues  in the case with shear and tidal force  in
LTA as well as  fluctuations.
\end{flushleft}

\newpage
\oddsidemargin -0.5in

\appendix
\section{The basic equations for the case with
fluctuations}
To be complete, here,  we   present a full
set of our basic equations, which have been used in our
renormalization analysis for Newtonian cosmology.
Those include fluctuations up to second order cumulants
as well as shear and tidal force.\\[1.5em]
{\bf (A) The RG equations for averaged variables}
\\[-.5em]
\begin{eqnarray}
{d\langle \rho \rangle \over d\ell }
&=& {{\cal S}} \bigg[\langle \theta \rangle -2 \bigg]
\langle
\rho \rangle
\label{re2rho}
\\[1em]
{d\langle \theta \rangle \over d\ell }
&=&{\cal S} \left[
 \left({1\over 3}\langle \theta \rangle -1\right)
  \langle \theta \rangle
   + 4\pi G \langle \rho \rangle
  + 6 \left(\langle \sigma_{+}\rangle^{2}
   + \langle \sigma_{-}\rangle^{2} \right)
- {2 \over 3} \langle {\theta }^2\rangle_{\rm c}
 + 6 \left(\langle \sigma_{+}^{2}\rangle_{\rm c}
   + \langle \sigma_{-}^{2}\rangle_{\rm c} \right)
\right]
\\[1em]
{d\langle \sigma_+\rangle \over d\ell }
&=&{{\cal S}} \left[
 \left({2 \over 3}\langle \theta \rangle -1\right)
\langle
\sigma_+\rangle
  - \langle \sigma_+\rangle^{2}
 + \langle \sigma_-\rangle^{2}  +\langle {E_+}\rangle
- {1 \over 3}\langle  \theta\sigma_+ \rangle_{\rm c}
  - \langle  \sigma_+^2\rangle_{\rm c}
 + \langle  \sigma_-^2\rangle_{\rm c}
\right]
\\[1em]
{d\langle \sigma_-\rangle \over d\ell }
&=&{{\cal S}} \left[
\left({2 \over 3}\langle \theta \rangle -1\right) \langle
\sigma_-\rangle
 +   2\langle \sigma_+\rangle
\langle\sigma_-\rangle +\langle E_-\rangle
 - {1 \over 3}\langle \theta  \sigma_-\rangle_{\rm c}
 +   2\langle \sigma_+ \sigma_-\rangle_{\rm c}
\right]
\\[1em]
{d\langle {E_+}\rangle \over d\ell }
&=& {{\cal S}} \bigg[
 \left(\langle \theta \rangle -2\right)  \langle
{E_+}\rangle
  + 4\pi G   \langle \rho \rangle \langle{{\sigma
}_+}\rangle
+ 4\pi G   \langle  \rho  {{\sigma }_+}\rangle_{\rm c}
 \bigg]
\\[1em]
{d\langle {E_-}\rangle \over d\ell }
&=&{{\cal S}} \bigg[
 \left(\langle \theta \rangle -2\right) \langle
{E_-}\rangle
 + 4\pi G  \langle \rho \rangle \langle {{\sigma
}_-}\rangle
+4\pi G  \langle  \rho  {{\sigma }_-}\rangle_{\rm c}
\bigg]
\end{eqnarray}
\vskip .5cm
{\bf (B) The RG equations for second order
cumulants}\\[-.5em]
\begin{eqnarray}
{d\langle {\rho }^2 \rangle_{\rm c} \over d\ell}
&=& 2 {\cal S} \bigg[ \left(\langle \theta \rangle
-2\right)
\langle  {\rho^2} \rangle_{\rm c}
+ \langle \rho \rangle  \langle   \rho \theta
\rangle_{\rm c}
  \bigg]
\\[1em]
{d\langle  {\theta }^2 \rangle_{\rm c} \over d \ell}
&=& 2{\cal S}\bigg[
 \left({2 \over 3} \langle
\theta \rangle - 1\right) \langle {\theta }^2
\rangle_{\rm c}  + 4\pi G  \langle   \rho \theta
\rangle_{\rm c}
 + 12 \left( \langle \sigma_+ \rangle  \langle
\theta
\sigma_+
\rangle_{\rm c}
 + \langle \sigma_- \rangle \langle  \theta
{\sigma }_{-}
\rangle_{\rm c} \right)
\bigg]
\\[1em]
{d\langle  \rho  \theta \rangle_{\rm c}  \over d\ell}
&=&  {\cal S}  \bigg[ \left(\frac{5}{3}\langle \theta
\rangle   - 3\right) \langle   \rho
\theta   \rangle_{\rm c}
    + 4\pi G \langle  {\rho }^2 \rangle_{\rm c}
+ \langle \rho \rangle  \langle  {\theta }^2
\rangle_{\rm c}
  + 12 \left( \langle \sigma_+ \rangle   \langle
\rho
\sigma_+
 \rangle_{\rm c} +\langle \sigma_- \rangle
 \langle  \rho   \sigma_-   \rangle_{\rm c}
 \right) \bigg]
\\[.5em]
{d\langle \rho   \sigma_+ \rangle_{\rm c}  \over d\ell}
&=& {\cal S} \bigg[ \left(\frac{5}{3}\langle \theta
\rangle  -2\langle \sigma_+ \rangle  - 3\right) \langle
\rho    \sigma_+ \rangle_{\rm c}
 + 2 \langle \sigma_-\rangle
 \langle \rho\sigma_-\rangle_{\rm c} + \frac{2}{3}
\langle  \sigma_+ \rangle \langle  \rho
\theta   \rangle_{\rm c}
 +  \langle \rho\rangle
\langle
\theta
\sigma_+ \rangle_{\rm c}
+  \langle \rho
{E_+}\rangle_{\rm c}
\bigg]
\\[1em]
{d\langle \rho   \sigma_- \rangle_{\rm c}   \over d\ell}
&=& {\cal S} \bigg[ \left(\frac{5}{3}\langle \theta
\rangle  +2\langle \sigma_+ \rangle  - 3\right) \langle
\rho    \sigma_- \rangle_{\rm c}
 + 2 \langle \sigma_-\rangle
 \langle \rho\sigma_+\rangle_{\rm c} + \frac{2}{3}
\langle  \sigma_- \rangle \langle  \rho
\theta
\rangle_{\rm c}
 +  \langle \rho\rangle
\langle
\theta
\sigma_- \rangle_{\rm c}
+  \langle \rho
{E_-}\rangle_{\rm c}
\bigg]
\\[1em]
{d\langle \rho   {E_+}\rangle_{\rm c}   \over d\ell}
&=&  {\cal S}  \bigg[
   2\left(\langle \theta \rangle -2\right)  \langle \rho
{E_+}\rangle_{\rm c}  +\langle {E_+}\rangle \langle \rho
\theta
\rangle_{\rm c}
  + \langle \rho \rangle   \langle \theta
{E_+}\rangle_{\rm c}
 +  4\pi G \left(\langle  \sigma_+ \rangle \langle {\rho
}^2
\rangle_{\rm c} + \langle \rho
\rangle   \langle
\rho    \sigma_+ \rangle_{\rm c} \right)
\bigg]
\\[1em]
{d\langle \rho   {E_-}\rangle_{\rm c}  \over d\ell}
&=&  {\cal S}  \bigg[
   2\left(\langle \theta \rangle -2\right)  \langle \rho
{E_-}\rangle_{\rm c}  +\langle {E_-}\rangle \langle \rho
\theta
\rangle_{\rm c}
  + \langle \rho \rangle   \langle \theta
{E_-}\rangle_{\rm c}
 +  4\pi G \left(\langle  \sigma_- \rangle \langle {\rho
}^2
\rangle_{\rm c} + \langle \rho
\rangle   \langle
\rho    \sigma_- \rangle_{\rm c} \right)
\bigg]
\\[3em]
{d\langle \theta   \sigma_+ \rangle_{\rm c} \over d\ell}
&=& {\cal S} \bigg[
2\left( {2 \over 3}\langle \theta \rangle -\langle
\sigma_+ \rangle - 1 \right) \langle
\theta
\sigma_+
\rangle_{\rm c}
+2\langle \sigma_-\rangle \langle
\theta \sigma_-\rangle_{\rm c}
+{2 \over 3} \langle \sigma_+\rangle \langle
\theta^2\rangle_{\rm c}
 +  \langle \theta   {E_+}\rangle_{\rm c}
 + 4\pi G   \langle \rho    \sigma_+ \rangle_{\rm c}
\nonumber
\\
& &
 + 12 \left( \langle  \sigma_+
\rangle
\langle  \sigma_+^2
\rangle_{\rm c} +  \langle  \sigma_-
\rangle
\langle  \sigma_+\sigma_-
\rangle_{\rm c}\right) \bigg]
\\[1em]
{d\langle \theta    \sigma_- \rangle_{\rm c} \over d\ell}
&=& {\cal S} \bigg[
2\left( {2 \over 3}\langle \theta \rangle +\langle
\sigma_+ \rangle - 1 \right) \langle
\theta
\sigma_-
\rangle_{\rm c}
+2\langle \sigma_-\rangle \langle
\theta \sigma_+\rangle_{\rm c}
+{2 \over 3} \langle \sigma_-\rangle \langle
\theta^2\rangle_{\rm c}
 +  \langle \theta   {E_-}\rangle_{\rm c}
 + 4\pi G   \langle \rho    \sigma_- \rangle_{\rm c}
\nonumber
\\
& &
 + 12 \left( \langle  \sigma_-
\rangle
\langle  \sigma_-^2
\rangle_{\rm c} +  \langle  \sigma_+
\rangle
\langle  \sigma_+\sigma_-
\rangle_{\rm c}\right) \bigg]
\\[1em]
{d\langle \theta   {E_+}\rangle_{\rm c}  \over d\ell}
&=& {\cal S} \bigg[\left(\frac{5}{3}\langle \theta
\rangle   - 3\right) \langle \theta {E_+}\rangle_{\rm c}
 + \langle {E_+}\rangle \langle  {{\theta
}^2}\rangle_{\rm c}
  + 4\pi G \left(  \langle \rho \rangle  \langle \theta
\sigma_+
\rangle_{\rm c} +\langle  \sigma_+ \rangle \langle \rho
\theta
\rangle_{\rm c}
   +    \langle \rho   {E_+}\rangle_{\rm c}
\right)
\nonumber
\\[.5em]
 & &
 + 12 \left( \langle  \sigma_+ \rangle  \langle
\sigma_+{E_+}
\rangle_{\rm c}  +  \langle  \sigma_- \rangle \langle
\sigma_-{E_+}
\rangle_{\rm c} \right)
\bigg]
\\[1em]
{d\langle \theta   {E_-}\rangle_{\rm c}  \over d\ell}
&=& {\cal S} \bigg[\left(\frac{5}{3}\langle \theta
\rangle   - 3\right) \langle \theta  {E_-}\rangle_{\rm c}
 + \langle {E_-}\rangle \langle  {{\theta
}^2}\rangle_{\rm c}
  + 4\pi G \left(  \langle \rho \rangle  \langle \theta
\sigma_-
\rangle_{\rm c} + \langle  \sigma_- \rangle \langle \rho
\theta
\rangle_{\rm c}
   +    \langle \rho   {E_-}\rangle_{\rm c}
\right)
\nonumber
\\[.5em]
 & &
 + 12 \left( \langle  \sigma_+ \rangle  \langle
\sigma_+{E_-}
\rangle_{\rm c}  +  \langle  \sigma_- \rangle  \langle
\sigma_-{E_-}
\rangle_{\rm c} \right)
\bigg]
\\[1em]
{d\langle \sigma_+^2\rangle_{\rm c}  \over d\ell}
&=& {\cal S} \bigg[
2\left( \frac{2}{3}\langle \theta \rangle
-2\langle  \sigma_+ \rangle -1 \right) \langle
\sigma_{+}^{2}\rangle_{\rm c}
 + \frac{4}{3} \langle  \sigma_+ \rangle \langle \theta
 \sigma_+ \rangle_{\rm c}
+  4 \langle  \sigma_- \rangle   \langle  \sigma_+
             \sigma_- \rangle_{\rm c}
 + 2  \langle \sigma_+  E_+ \rangle_{\rm c}
\bigg]
\\[1em]
{d\langle \sigma_-^2\rangle_{\rm c}   \over d\ell}
&=&  {\cal S} \bigg[
2\left( \frac{2}{3}\langle \theta \rangle
+2\langle  \sigma_+ \rangle -1 \right) \langle
\sigma_{-}^{2}\rangle_{\rm c}
 + \frac{4}{3} \langle  \sigma_- \rangle \langle \theta
 \sigma_- \rangle_{\rm c}
 +  4 \langle  \sigma_- \rangle   \langle  \sigma_+
             \sigma_- \rangle_{\rm c}
 + 2  \langle \sigma_-  E_- \rangle_{\rm c}
\bigg]
\\[1em]
{d\langle  \sigma_+\sigma_- \rangle_{\rm c} \over d\ell}
&=&  {\cal S} \bigg[
2\left( \frac{2}{3}\langle \theta \rangle
-1 \right) \langle
\sigma_{+}\sigma_{-}\rangle_{\rm c}
 + \frac{2}{3} \left( \langle  \sigma_+ \rangle \langle
\theta
 \sigma_- \rangle_{\rm c} + \langle  \sigma_- \rangle
\langle \theta
 \sigma_+ \rangle_{\rm c} \right)
 +  2 \langle
\sigma_- \rangle \left(  \langle  \sigma_+^2
\rangle_{\rm c} +\langle
\sigma_-^2 \rangle_{\rm c}\right)
\nonumber
\\[.5em]
 & &
 +  \langle \sigma_+  E_- \rangle_{\rm c}
 +  \langle \sigma_-  E_+ \rangle_{\rm c}
\bigg]
\\[1em]
{d\langle E_{+}^{2}\rangle_{\rm c}  \over d\ell}
&=& 2 {\cal S}  \bigg[
 \left(\langle \theta \rangle -2)\langle E_{+}^{2}
\rangle_{\rm c}
 + \langle {E_+}\rangle   \langle \theta
{E_+}\rangle_{\rm c}
+ 4\pi G (\langle \rho \rangle  \langle \sigma_+  {E_+}
\rangle_{\rm c} +\langle  \sigma_+ \rangle \langle \rho
{E_+}\rangle_{\rm c}
\right)
\bigg]
\\[1em]
{d\langle E_{-}^{2}\rangle_{\rm c}  \over d\ell}
&=& 2 {\cal S}  \bigg[
 \left(\langle \theta \rangle -2)\langle E_{-}^{2}
\rangle_{\rm c}
 + \langle {E_-}\rangle   \langle \theta
{E_-}\rangle_{\rm c}
+ 4\pi G (\langle \rho \rangle  \langle \sigma_-  {E_-}
\rangle_{\rm c} + \langle  \sigma_- \rangle \langle \rho
{E_-}\rangle_{\rm c}
\right)
\bigg]
\\[1em]
{d\langle {E_+} {E_-}\rangle_{\rm c}  \over d\ell}
&=&  {\cal S}\bigg[
 2\left(\langle \theta \rangle -2\right)\langle
E_{+}E_{-}
\rangle_{\rm c}
 + \langle {E_-}\rangle   \langle \theta
{E_+}\rangle_{\rm c}+ \langle {E_+}\rangle \langle
\theta   {E_-}\rangle_{\rm c}
+ 4\pi G \langle \rho \rangle \left( \langle \sigma_+
{E_-}
\rangle_{\rm c}+\langle \sigma_-  {E_+}
\rangle_{\rm c}\right)
\nonumber
\\[.5em]
 & &
+ 4\pi G \left( \langle  \sigma_+ \rangle \langle \rho
{E_-}\rangle_{\rm c}  +
\langle  \sigma_- \rangle \langle \rho
{E_+}\rangle_{\rm c} \right)
\bigg]
\\[1em]
{d\langle\sigma_+  {E_+}  \rangle_{\rm c} \over d\ell}
&=& {\cal S}  \bigg[
\left(\frac{5}{3}\langle \theta \rangle -2 \langle
\sigma_+
 \rangle - 3\right) \langle \sigma_+ {E_+} \rangle_{\rm c}
 + 2  \langle  \sigma_- \rangle   \langle
                  \sigma_- E_+\rangle_{\rm c}
+\langle {E_+}\rangle  \langle \theta\sigma_+
\rangle_{\rm c}
+ \langle E_{+}^{2}\rangle_{\rm c}
\nonumber
\\[.5em]
 & &
 + \frac{2}{3}\langle  \sigma_+ \rangle \langle \theta
{E_+}\rangle_{\rm c}
 + 4\pi G \left( \langle \rho \rangle  \langle
\sigma_+^2\rangle_{\rm c}
+\langle \sigma_+ \rangle  \langle
\rho\sigma_+\rangle_{\rm c}\right)
\bigg]
\\[3em]
{d\langle \sigma_+ E_-  \rangle_{\rm c}  \over d\ell}
&=& {\cal S}  \bigg[
\left(\frac{5}{3}\langle \theta \rangle -2 \langle
\sigma_+\rangle - 3\right) \langle \sigma_+ {E_-}
\rangle_{\rm c}
 + 2  \langle  \sigma_- \rangle   \langle
                  \sigma_- E_-\rangle_{\rm c}
+\langle {E_-}\rangle  \langle \theta\sigma_+
\rangle_{\rm c}
+ \langle E_{+}E_{-}\rangle_{\rm c}
\nonumber
\\[.5em]
 & &
 + \frac{2}{3}\langle  \sigma_+ \rangle \langle \theta
{E_-}\rangle_{\rm c}
 + 4\pi G \left( \langle \rho \rangle  \langle
\sigma_+\sigma_-\rangle_{\rm c}
+\langle \sigma_- \rangle  \langle
\rho\sigma_+\rangle_{\rm c}\right)
\bigg]
\\[1em]
{d\langle \sigma_- {E_+}  \rangle_{\rm c}  \over d\ell}
&=& {\cal S}  \bigg[
\left(\frac{5}{3}\langle \theta \rangle + 2 \langle
\sigma_+\rangle - 3\right) \langle \sigma_- {E_+}
\rangle_{\rm c}
 + 2  \langle  \sigma_- \rangle   \langle
                  \sigma_+ E_+\rangle_{\rm c}
+\langle {E_+}\rangle  \langle \theta\sigma_-
\rangle_{\rm c}
+ \langle E_{+}E_{-}\rangle_{\rm c}
\nonumber
\\[.5em]
 & &
 + \frac{2}{3}\langle  \sigma_- \rangle \langle \theta
{E_+}\rangle_{\rm c}
 + 4\pi G \left( \langle \rho \rangle  \langle
\sigma_+\sigma_-\rangle_{\rm c}
+\langle \sigma_+ \rangle  \langle
\rho\sigma_-\rangle_{\rm c}\right)
\bigg]
\\[1em]
{d\langle \sigma_- {E_-}  \rangle_{\rm c}  \over d\ell}
&=& {\cal S}  \bigg[
\left(\frac{5}{3}\langle \theta \rangle + 2 \langle
\sigma_+\rangle - 3\right) \langle \sigma_- {E_-}
\rangle_{\rm c}
 + 2  \langle  \sigma_- \rangle   \langle
                  \sigma_+ E_-\rangle_{\rm c}
+\langle {E_-}\rangle  \langle \theta\sigma_-
\rangle_{\rm c}
+ \langle E_{-}^{2}\rangle_{\rm c}
\nonumber
\\[.5em]
 & &
 + \frac{2}{3}\langle  \sigma_- \rangle \langle \theta
{E_-}\rangle_{\rm c}
 + 4\pi G \left( \langle \rho \rangle  \langle
\sigma_-^2\rangle_{\rm c}
+\langle \sigma_- \rangle  \langle
\rho\sigma_-\rangle_{\rm c}\right)
\bigg]
\label{re2sig-e+}
\end{eqnarray}

\newpage

\begin{figure}[h]
\caption[fig.1]{The RG flow  toward larger scale on the
$H$-$\Omega$ plane
in the case with neither shear
nor fluctuations.  Here we define  the
Hubble parameter $H$ and the density parameter
$\Omega$  as $H\equiv \la\theta\ra /3$ and $\Omega\equiv
8\pi G \la\rho\ra / 3H^2$, respectively. The three
important fixed  points  exist:
(1) fixed point Q (Quiescent Universe),
(2) fixed point  E (``Einstein-de Sitter" Universe)
(3) fixed point M (``Milne" Universe).
The RG flow  converges to the fixed
point Q  or escapes to infinity.
The fixed point seems to be a line in the $H$-$\Omega$
plane, but it is a point that all physical variables
vanish in the ``phase" space of $\la F\ra=(\la\rho\ra ,
\la\theta\ra)$. $\Omega$ turns out to be some finite
value, which depends on the limit of $\la\rho\ra \rightarrow
0$ and $
\la\theta\ra\rightarrow 0$}
\vspace{1cm}

\caption[fig.2]{The scale dependence  of (a)
$\log\Omega$, (b)
$H$, (c) $\Delta_\rho$ and (d) $\Delta_\theta$  in shear-free 
case with $s=0.5$ and $\Delta_0
= ~(1) ~ 10^{-1}, ~(2) ~ 10^{-1.5}, ~(3) ~ 10^{-2},~ (4)
~ 10^{-2.5}$. We solve the RG flows  from the fixed
point E  toward smaller scale.
$L\equiv\exp{\ell}$ represents the ratio of
the scale of averaged region to the horizon scale.
Every
flow monotonically converges to M as  $H$ becomes larger
($H
\nearrow$) and $\Omega$ gets  smaller  ($\Omega
\searrow$) toward smaller scale.   The critical scale
where  $\Omega$ and
$H$ deviate from the EdS values  depends on the
magnitude of
$\Delta_0$.}
\vspace{1cm}

\caption[fig.3]{The $s$
dependence of the RG flows of (a) $\log\Omega$,  (b) $H$,
(c) $\Delta_\rho$ and (d) $\Delta_\theta$  from the
fixed  point E at horizon scale in shear-free case.
Here we depict the cases of 
$\Delta_0=10^{-2}$ and $s =~ (1)
~ 0.3,~ (2) ~ 0.5,~ (3) ~ 0.7,~ (4) ~ 0.9$.
$\Omega$ begins to decline
beyond a critical scale, a few decades smaller than
horizon scale, which depends on the magnitude of
$\Delta_0$.
The slope at smaller scale is determined
by $\lambda^{[2]}_{\rm min}=s/(1-s)$ at the fixed point
M. The slope for the case $(2)$ is equal to 1.
$H$ also changes from the value at E to the value at M
beyond the above critical scale.
As the value of $s$ increases,
the critical scale shifts toward smaller scale.}
\vspace{1cm}

\caption[fig.4]{The RG flow
 toward smaller scale on the $H$-$\Omega$ plane in the
case that
$s=0.5$ and $\bar{\Delta}_{\sigma_+,0}  =
\bar{\Delta}_{E_+,0} = 10^{-2}$.
As for the density perturbations, we choose
$\Delta_0
=~(1)~10^{-0.75},~(2)~10^{-1},~(3)~10^{-1.25},
~(4)~10^{-1.5}$. If both 
$\bar{\Delta}_{\sigma_+,0}/\Delta_0$ and
$\bar{\Delta}_{E_+,0}/\Delta_0$
are less than
$R_{cr} \approx 0.1495$ ((1), (2)),
every flow converges to M as $H$ increases ($H \nearrow$)
to 1 and
$\Omega$ decreases ($\Omega
\searrow$) to 0 toward smaller scales. On the other hand,
if $\bar{\Delta}_{\sigma_+,0}$  or
$\bar{\Delta}_{E_+,0}$ are larger than  the  above
critical value $R_{cr}$ ((3), (4)), the flow heading for M comes to
turn  around and go away to infinity.}
\vspace{1cm}

\caption[fig.5]{The RG flow of (a) $\log\Omega$,  (b)
$H$, (c) $\log(\la\sigma_\pm\ra/H)$ and 
(d)$\log( \la E_\pm\ra/H^2)$ from
the fixed point E toward smaller scale in the case that
$s=0.5$,  $\bar{\Delta}_{\sigma_+,0}  =
\bar{\Delta}_{E_+,0} = 10^{-2}$ and $\Delta_0
=~(1)~10^{-0.75},~(2)~10^{-1},~(3)~10^{-1.25},
~(4)~10^{-1.5}$. In the case of (1) or (2)
($\bar{\Delta}_{\sigma_+,0}/\Delta_0, 
\bar{\Delta}_{E_+,0}/\Delta_0 \lsim
R_{cr} \approx 0.1495$), 
the RG flow   converges to M toward smaller
scales as the case without shear and tidal force (see Fig.
2). In the case of (3) or (4)
($\bar{\Delta}_{\sigma_+,0}/\Delta_0,
\bar{\Delta}_{E_+,0}/\Delta_0
\gsim R_{cr}$), 
however, the RG flow does not converge to
M but rather diverges to infinity.
 We also depicted the flow of fluctuations;
(e)
$\log \Delta_\rho $, (f) $\log\Delta_\theta $,
(g) $\log\bar{\Delta}{\sigma_+} $ and (h) 
$\log\bar{\Delta}_{E_+} $ for the same initial values as the above. 
In the case of (1) or (2), all fluctuations but $\Delta_\rho$
damp off as the RG flow converges to M, while
$\Delta_\rho$ approaches the finite value. 
In the case of (3) or (4), all of them 
diverges to infinity. }
\vspace{1cm}

\caption[fig.6]{The $s$ dependence
of the RG flows from fixed point E of (a) $\log\Omega$,
(b) $H$, (c) $\log(\la\sigma_\pm\ra/H)$ and 
(d)$\log( \la E_\pm\ra/H^2)$.
Here we depict the cases of $s=~ (1)
~ 0.3,~ (2) ~ 0.5,~ (3) ~ 0.7, ~ (4) ~ 0.9$ with the
boundary conditions,
$\Delta_0 =10^{-1.5}$ and
$\bar{\Delta}_{\sigma_+, 0} =\bar{\Delta}_{E_+,0} = ~
10^{-2.5}$ at horizon scale.
The results for  $\log\Omega$
and
$H$ are quite similar to Fig. 3, if the flow
converges to the fixed point M. The slopes of
$\la\sigma_\pm\ra$ and $\la E_\pm\ra$
 at larger and smaller scale are determined by
$\lambda^{[2]}_{\rm min}$ at  fixed points E and M, i.e.
$-4s/(3-2s)$ and $2s/(1-s)$, respectively. As
$s$  increases, the turning point of the slope shifts
toward smaller scale  and the slope
becomes less steep. We also depicted the flow 
of fluctuations;
(e)
$\log \Delta_\rho $, (f) $\log\Delta_\theta $,
(g) $\log\bar{\Delta}_{\sigma_+} $ and (h) 
$\log\bar{\Delta}_{E_+} $ for the same range of $s$.
$\Delta_0=10^{-1.5}$ and
$\bar{\Delta}_{\sigma_+, 0} =\bar{\Delta}_{E_+, 0}= ~
10^{-2.5}$ at  horizon scale in the case of LTA. We find
that
$\Delta_\rho^2 $ = constant near the fixed point M, while
it decreases linearly near the fixed point E, which
behavior is determined by the eigenvalue $-4s/(3-2s)$.}
\vspace{1cm}

\caption[fig.7]{The scale dependence
of (a) $\log \la\rho\ra$, (b) $H$, 
(c) $\log \Delta_{\theta}$ and
(d) $\log \Delta_{\rho}$
beyond horizon scale
for $s=0.5$
with the boundary conditions,
$\Delta_0=10^{-2.5}$ at horizon scale in shear-free case.  
In the flow 
departing from the  fixed point E, $\la\rho\ra$
approaches zero toward both smaller and larger
scales, for which the slopes at
larger and smaller scale are
determined by $\lambda^{[2]}_{\rm min}$ at fixed points,
Q and M, i.e.
$-2s$ and $2s/(1-2s)$, respectively.
$H$ also vanishes at larger scale,
after it becomes negative around $L=10^5$, which causes
the divergence of $\Delta_{\theta}$ around the scale. 
The slope of both $\Delta_{\rho}$ and $\Delta_{\theta}$ around
the fixed point E are
determined by $\lambda^{[2]}_{\rm min}=-4s/(3-2s)$ toward 
smaller scale and $\lambda^{[2]}_{\rm max}=6s/(3-2s)$ 
toward larger scale at the point.}
\label{beyondscale.fig}
\end{figure}


\begin{thebibliography}{99}
\bibitem{foot1}
In the present
consideration, we exclude the direct observations of
$\Omega$ based on the space-time geometry.
\bibitem{hamilton-foot}In this approximation,
the third and higher  order spatial derivatives are
systematically  neglected in the equation of motion
for the tidal force.  This is a generalization of
the Zel'dovich approximation, and  accuracies of
these local approximations are examined for
spheroidal configurations by Hui and Bertschinger
\cite{hui96} .  It is also worth noticing that the
choice of  approximation scheme does not change our
basic idea in this paper and many of them give
similar results.
\bibitem{foot3}
The  RG equations are written in terms
of a scale parameter
$\ell$.  Then a ``stability" here does not mean
dynamical stability. It means  what kind of asymptotic
behavior of some averaged variables would be expected
when  a scale is changed.
\bibitem{foot4}
Other choices of $s$ are mathematically
viable, however they seem  not to provide us
any interesting  model of the Universe.
\bibitem{foot5}
Such a similarity solution with $s\neq 1$ is related
to ``partial similarity"\cite{tomita} or 
"kinematic self-similarity"\cite{coley} in general
relativity.
\bibitem{foot6}
If the structure of the Universe is
described by Rayleigh-L\'{e}vy fractal, the higher-order
cumulants are given as the three-point correlation
function $\zeta_{123} = {1
\over 2}
(\xi_{12}\xi_{23}+\xi_{23}\xi_{31}+\xi_{31}\xi_{12})$
and the
four-point correlation function
$\eta_{1234} = {1 \over 4}
(\xi_{12}\xi_{23}\xi_{34} +\cdots {\rm 12 terms})$
\bibitem{foot7}
This relation may be true only for the present case
without higher-order cumulants.
\bibitem{foot8}
In this approximation, the
magnetic part  of the  Weyl tensor  is neglected.
\bibitem{foot9}
Here, ``initial" means
$\ell =0$, which denotes the largest averaging scale of
our Universe ($\sim$ horizon scale)
\end{thebibliography}
\end{document}